\input harvmac.tex
\input epsf.tex
\input amssym
\input ulem.sty
\input graphicx.tex

%\draftmode

\let\includefigures=\iftrue
\let\useblackboard=\iftrue
\newfam\black

%Figure Stuff
\includefigures
\message{If you do not have epsf.tex (to include figures),}
\message{change the option at the top of the tex file.}
\input epsf
\def\figin{\epsfcheck\figin}\def\figins{\epsfcheck\figins}
\def\epsfcheck{\ifx\epsfbox\UnDeFiNeD
\message{(NO epsf.tex, FIGURES WILL BE IGNORED)}
\gdef\figin##1{\vskip2in}\gdef\figins##1{\hskip.5in}% blank space instead
\else\message{(FIGURES WILL BE INCLUDED)}%
\gdef\figin##1{##1}\gdef\figins##1{##1}\fi}
\def\DefWarn#1{}
\def\figinsert{\goodbreak\midinsert}
\def\ifig#1#2#3{\DefWarn#1\xdef#1{fig.~\the\figno}
\writedef{#1\leftbracket fig.\noexpand~\the\figno}%
\figinsert\figin{\centerline{#3}}\medskip\centerline{\vbox{
\baselineskip12pt\advance\hsize by -1truein
\noindent\footnotefont{\bf Fig.~\the\figno:} #2}}
%\bigskip
\endinsert\global\advance\figno by1}
%%%
\else
\def\ifig#1#2#3{\xdef#1{fig.~\the\figno}
\writedef{#1\leftbracket fig.\noexpand~\the\figno}%
%\figinsert\figin{\centerline{#3}}\medskip
%\centerline{\vbox{\baselineskip12pt
%\advance\hsize by -1truein\noindent
%\footnotefont{\bf Fig.~\the\figno:} #2}}
%\bigskip\endinsert
\global\advance\figno by1} \fi

\def \la {\langle}
\def \ra {\rangle}

\def \pa {\partial}

\catcode`\@=11
\def\slash#1{\mathord{\mathpalette\c@ncel{#1}}}
\overfullrule=0pt
\def\AA{{\cal A}}

\def\underrel#1\over#2{\mathrel{\mathop{\kern\z@#1}\limits_{#2}}}

\catcode`\@=12

%%%%%%%%%%%%%%%%%%%%%%%%%%%%%%%%%%%%%%%%%%%%%%%%%%%%%%%%%%%%%%

%%%%%%%%%%%%%%%%%%%%%%%%%%%%%%%%%%%%%%%%%%%%%%%%%%%%%%%%%%%%%%
% new defs:

%\def\ra{{\rightarrow}}

%\Simmons-DuffinWLQ
\lref\SimmonsDuffinWLQ{
  D.~Simmons-Duffin,
  ``The Lightcone Bootstrap and the Spectrum of the 3d Ising CFT,''
JHEP {\bf 1703}, 086 (2017).
[arXiv:1612.08471 [hep-th]].
%%CITATION = arXiv:1612.08471%%
}

%\Caron-HuotVEP
\lref\Simon{
  S.~Caron-Huot,
  ``Analyticity in Spin in Conformal Theories,''
JHEP {\bf 1709}, 078 (2017).
[arXiv:1703.00278 [hep-th]].
%%CITATION = arXiv:1703.00278%%
}

%\HeemskerkPN
\lref\HPPS{
  I.~Heemskerk, J.~Penedones, J.~Polchinski and J.~Sully,
  ``Holography from Conformal Field Theory,''
JHEP {\bf 0910}, 079 (2009).
[arXiv:0907.0151 [hep-th]].
%%CITATION = arXiv:0907.0151%%
}

%\GiombiKC
\lref\GiombiKC{
  S.~Giombi, S.~Minwalla, S.~Prakash, S.~P.~Trivedi, S.~R.~Wadia and X.~Yin,
  ``Chern-Simons Theory with Vector Fermion Matter,''
Eur.\ Phys.\ J.\ C {\bf 72}, 2112 (2012).
[arXiv:1110.4386 [hep-th]].
%%CITATION = arXiv:1110.4386%%
}

%\AharonyNH
\lref\AharonyNH{
  O.~Aharony, G.~Gur-Ari and R.~Yacoby,
  ``Correlation Functions of Large N Chern-Simons-Matter Theories and Bosonization in Three Dimensions,''
JHEP {\bf 1212}, 028 (2012).
[arXiv:1207.4593 [hep-th]].
%%CITATION = WIS-13-12-JUL-DPPA%%
}

%\GurAriIS
\lref\GurAriIS{
  G.~Gur-Ari and R.~Yacoby,
  ``Correlators of Large N Fermionic Chern-Simons Vector Models,''
JHEP {\bf 1302}, 150 (2013).
[arXiv:1211.1866 [hep-th]].
%%CITATION = arXiv:1211.1866%%
}

%\JainNZA
\lref\Minwalla{
  S.~Jain, M.~Mandlik, S.~Minwalla, T.~Takimi, S.~R.~Wadia and S.~Yokoyama,
  ``Unitarity, Crossing Symmetry and Duality of the S-matrix in large N Chern-Simons theories with fundamental matter,''
JHEP {\bf 1504}, 129 (2015).
[arXiv:1404.6373 [hep-th]].
%%CITATION = TIFR-TH-14-12%%
}

%\BedhotiyaUGA
\lref\PRAKASH{
  A.~Bedhotiya and S.~Prakash,
  ``A test of bosonization at the level of four-point functions in Chern-Simons vector models,''
JHEP {\bf 1512}, 032 (2015).
[arXiv:1506.05412 [hep-th]].
%%CITATION = arXiv:1506.05412%%
}

%\LeonhardtDU
\lref\LeonhardtDU{
  T.~Leonhardt and W.~Ruhl,
  ``The Minimal conformal O(N) vector sigma model at d = 3,''
J.\ Phys.\ A {\bf 37}, 1403 (2004).
[hep-th/0308111].
%%CITATION = hep-th/0308111%%
}

%\MaldacenaWAA
\lref\MaldacenaWAA{
  J.~Maldacena, S.~H.~Shenker and D.~Stanford,
  ``A bound on chaos,''
JHEP {\bf 1608}, 106 (2016).
[arXiv:1503.01409 [hep-th]].
%%CITATION = arXiv:1503.01409%%
}

%\PenedonesUE
\lref\PenedonesUE{
  J.~Penedones,
  ``Writing CFT correlation functions as AdS scattering amplitudes,''
JHEP {\bf 1103}, 025 (2011).
[arXiv:1011.1485 [hep-th]].
%%CITATION = arXiv:1011.1485%%
}

%\Simmons-DuffinNUB
\lref\SimmonsDuffinNUB{
  D.~Simmons-Duffin, D.~Stanford and E.~Witten,
  ``A spacetime derivation of the Lorentzian OPE inversion formula,''
[arXiv:1711.03816 [hep-th]].
%%CITATION = arXiv:1711.03816%%
}

%\SezginPT
\lref\SezginPT{
  E.~Sezgin and P.~Sundell,
  ``Holography in 4D (super) higher spin theories and a test via cubic scalar couplings,''
JHEP {\bf 0507}, 044 (2005).
[hep-th/0305040].
%%CITATION = hep-th/0305040%%
}

%\GurAriIS
\lref\GurAriIS{
  G.~Gur-Ari and R.~Yacoby,
  ``Correlators of Large N Fermionic Chern-Simons Vector Models,''
JHEP {\bf 1302}, 150 (2013).
[arXiv:1211.1866 [hep-th]].
%%CITATION = arXiv:1211.1866%%
}

%\DolanUW
\lref\DolanUW{
  F.~A.~Dolan and H.~Osborn,
  ``Implications of N=1 superconformal symmetry for chiral fields,''
Nucl.\ Phys.\ B {\bf 593}, 599 (2001).
[hep-th/0006098].
%%CITATION = hep-th/0006098%%
}

%\DolanIY
\lref\DolanIY{
  F.~A.~Dolan and H.~Osborn,
  ``Conformal partial wave expansions for N=4 chiral four point functions,''
Annals Phys.\  {\bf 321}, 581 (2006).
[hep-th/0412335].
%%CITATION = hep-th/0412335%%
}

%\DolanUT
\lref\DolanUT{
  F.~A.~Dolan and H.~Osborn,
  ``Conformal four point functions and the operator product expansion,''
Nucl.\ Phys.\ B {\bf 599}, 459 (2001).
[hep-th/0011040].
%%CITATION = hep-th/0011040%%
}

%\AldayVKK
\lref\AldayVKK{
  L.~F.~Alday and S.~Caron-Huot,
  ``Gravitational S-matrix from CFT dispersion relations,''
[arXiv:1711.02031 [hep-th]].
%%CITATION = arXiv:1711.02031%%
}

%\MaldacenaJN
\lref\MaldacenaJN{
  J.~Maldacena and A.~Zhiboedov,
  ``Constraining Conformal Field Theories with A Higher Spin Symmetry,''
J.\ Phys.\ A {\bf 46}, 214011 (2013).
[arXiv:1112.1016 [hep-th]].
%%CITATION = arXiv:1112.1016%%
}

%\MaldacenaSF
\lref\MaldacenaSF{
  J.~Maldacena and A.~Zhiboedov,
  ``Constraining conformal field theories with a slightly broken higher spin symmetry,''
Class.\ Quant.\ Grav.\  {\bf 30}, 104003 (2013).
[arXiv:1204.3882 [hep-th]].
%%CITATION = PUPT-2410%%
}

%\MaldacenaNZ
\lref\MaldacenaNZ{
  J.~M.~Maldacena and G.~L.~Pimentel,
  ``On graviton non-Gaussianities during inflation,''
JHEP {\bf 1109}, 045 (2011).
[arXiv:1104.2846 [hep-th]].
%%CITATION = arXiv:1104.2846%%
}

%\BzowskiSZA
\lref\BzowskiSZA{
  A.~Bzowski, P.~McFadden and K.~Skenderis,
  ``Implications of conformal invariance in momentum space,''
JHEP {\bf 1403}, 111 (2014).
[arXiv:1304.7760 [hep-th]].
%%CITATION = arXiv:1304.7760%%
}

%\BekaertCEA
\lref\BekaertCEA{
  X.~Bekaert, J.~Erdmenger, D.~Ponomarev and C.~Sleight,
  ``Towards holographic higher-spin interactions: Four-point functions and higher-spin exchange,''
JHEP {\bf 1503}, 170 (2015).
[arXiv:1412.0016 [hep-th]].
%%CITATION = arXiv:1412.0016%%
}

%\BekaertTVA
\lref\BekaertTVA{
  X.~Bekaert, J.~Erdmenger, D.~Ponomarev and C.~Sleight,
  ``Quartic AdS Interactions in Higher-Spin Gravity from Conformal Field Theory,''
JHEP {\bf 1511}, 149 (2015).
[arXiv:1508.04292 [hep-th]].
%%CITATION = MPP-2015-178%%
}

%\BekaertEZC
\lref\BekaertEZC{
  X.~Bekaert, J.~Erdmenger, D.~Ponomarev and C.~Sleight,
  ``Bulk quartic vertices from boundary four-point correlators,''
[arXiv:1602.08570 [hep-th]].
%%CITATION = arXiv:1602.08570%%
}

%\SleightDBA
\lref\SleightDBA{
  C.~Sleight and M.~Taronna,
  ``Higher Spin Interactions from Conformal Field Theory: The Complete Cubic Couplings,''
Phys.\ Rev.\ Lett.\  {\bf 116}, no. 18, 181602 (2016).
[arXiv:1603.00022 [hep-th]].
%%CITATION = MPP-2016-25%%
}

%\SleightCAX
\lref\SleightCAX{
  C.~Sleight and M.~Taronna,
  ``Feynman rules for higher-spin gauge fields on AdS$_{d+1}$,''
JHEP {\bf 1801}, 060 (2018).
[arXiv:1708.08668 [hep-th]].
%%CITATION = arXiv:1708.08668%%
}

%\KlebanovJA
\lref\KlebanovJA{
  I.~R.~Klebanov and A.~M.~Polyakov,
  ``AdS dual of the critical O(N) vector model,''
Phys.\ Lett.\ B {\bf 550}, 213 (2002).
[hep-th/0210114].
%%CITATION = hep-th/0210114%%
}

%\GiombiWH
\lref\GiombiWH{
  S.~Giombi and X.~Yin,
  ``Higher Spin Gauge Theory and Holography: The Three-Point Functions,''
JHEP {\bf 1009}, 115 (2010).
[arXiv:0912.3462 [hep-th]].
%%CITATION = arXiv:0912.3462%%
}

%\VasilievEV
\lref\VasilievEV{
  M.~A.~Vasiliev,
  ``Nonlinear equations for symmetric massless higher spin fields in (A)dS(d),''
Phys.\ Lett.\ B {\bf 567}, 139 (2003).
[hep-th/0304049].
%%CITATION = hep-th/0304049%%
}

%\AharonyNS
\lref\AharonyNS{
  O.~Aharony, S.~Giombi, G.~Gur-Ari, J.~Maldacena and R.~Yacoby,
  ``The Thermal Free Energy in Large N Chern-Simons-Matter Theories,''
JHEP {\bf 1303}, 121 (2013).
[arXiv:1211.4843 [hep-th]].
%%CITATION = WIS-18-12-NOV-DPPA%%
}

%\AharonyNH
\lref\AharonyNH{
  O.~Aharony, G.~Gur-Ari and R.~Yacoby,
  ``Correlation Functions of Large N Chern-Simons-Matter Theories and Bosonization in Three Dimensions,''
JHEP {\bf 1212}, 028 (2012).
[arXiv:1207.4593 [hep-th]].
%%CITATION = WIS-13-12-JUL-DPPA%%
}

%\AharonyJZ
\lref\AharonyJZ{
  O.~Aharony, G.~Gur-Ari and R.~Yacoby,
  ``d=3 Bosonic Vector Models Coupled to Chern-Simons Gauge Theories,''
JHEP {\bf 1203}, 037 (2012).
[arXiv:1110.4382 [hep-th]].
%%CITATION = arXiv:1110.4382%%
}

%\GurAriIS
\lref\GurAriIS{
  G.~Gur-Ari and R.~Yacoby,
  ``Correlators of Large N Fermionic Chern-Simons Vector Models,''
JHEP {\bf 1302}, 150 (2013).
[arXiv:1211.1866 [hep-th]].
%%CITATION = arXiv:1211.1866%%
}

%\SezginRT
\lref\SezginRT{
  E.~Sezgin and P.~Sundell,
  ``Massless higher spins and holography,''
Nucl.\ Phys.\ B {\bf 644}, 303 (2002), Erratum: [Nucl.\ Phys.\ B {\bf 660}, 403 (2003)].
[hep-th/0205131].
%%CITATION = hep-th/0205131%%
}

%\DidenkoTV
\lref\DidenkoTV{
  V.~E.~Didenko and E.~D.~Skvortsov,
  ``Exact higher-spin symmetry in CFT: all correlators in unbroken Vasiliev theory,''
JHEP {\bf 1304}, 158 (2013).
[arXiv:1210.7963 [hep-th]].
%%CITATION = arXiv:1210.7963%%
}

%\GiombiHKJ
\lref\GiombiHKJ{
  S.~Giombi and V.~Kirilin,
  ``Anomalous dimensions in CFT with weakly broken higher spin symmetry,''
JHEP {\bf 1611}, 068 (2016).
[arXiv:1601.01310 [hep-th]].
%%CITATION = PUPT-2495%%
}

%\GiombiZWA
\lref\GiombiZWA{
  S.~Giombi, V.~Gurucharan, V.~Kirilin, S.~Prakash and E.~Skvortsov,
  ``On the Higher-Spin Spectrum in Large N Chern-Simons Vector Models,''
JHEP {\bf 1701}, 058 (2017).
[arXiv:1610.08472 [hep-th]].
%%CITATION = PUPT-2512%%
}

%\GiombiRHM
\lref\GiombiRHM{
  S.~Giombi, V.~Kirilin and E.~Skvortsov,
  ``Notes on Spinning Operators in Fermionic CFT,''
JHEP {\bf 1705}, 041 (2017).
[arXiv:1701.06997 [hep-th]].
%%CITATION = PUPT-2517%%
}

%\GiombiKC
\lref\GiombiKC{
  S.~Giombi, S.~Minwalla, S.~Prakash, S.~P.~Trivedi, S.~R.~Wadia and X.~Yin,
  ``Chern-Simons Theory with Vector Fermion Matter,''
Eur.\ Phys.\ J.\ C {\bf 72}, 2112 (2012).
[arXiv:1110.4386 [hep-th]].
%%CITATION = arXiv:1110.4386%%
}

%\GiombiRZ
\lref\GiombiRZ{
  S.~Giombi, S.~Prakash and X.~Yin,
  ``A Note on CFT Correlators in Three Dimensions,''
JHEP {\bf 1307}, 105 (2013).
[arXiv:1104.4317 [hep-th]].
%%CITATION = arXiv:1104.4317%%
}

%\JainNZA
\lref\JainNZA{
  S.~Jain, M.~Mandlik, S.~Minwalla, T.~Takimi, S.~R.~Wadia and S.~Yokoyama,
  ``Unitarity, Crossing Symmetry and Duality of the S-matrix in large N Chern-Simons theories with fundamental matter,''
JHEP {\bf 1504}, 129 (2015).
[arXiv:1404.6373 [hep-th]].
%%CITATION = TIFR-TH-14-12%%
}

%\Arkani-HamedBZA
\lref\ArkaniHamedBZA{
  N.~Arkani-Hamed and J.~Maldacena,
  ``Cosmological Collider Physics,''
[arXiv:1503.08043 [hep-th]].
%%CITATION = arXiv:1503.08043%%
}

%\VasilievBA
\lref\VasilievBA{
  M.~A.~Vasiliev,
  ``Higher spin gauge theories: Star product and AdS space,''
In *Shifman, M.A. (ed.): The many faces of the superworld* 533-610.
[hep-th/9910096].
%%CITATION = FIAN-TD-24-99%%
}

%\VasilievEN
\lref\VasilievEN{
  M.~A.~Vasiliev,
  ``Consistent equation for interacting gauge fields of all spins in (3+1)-dimensions,''
Phys.\ Lett.\ B {\bf 243}, 378 (1990)..
%%CITATION = LEBEDEV-90-29%%
}

%\GiombiVG
\lref\GiombiVG{
  S.~Giombi and X.~Yin,
  ``Higher Spins in AdS and Twistorial Holography,''
JHEP {\bf 1104}, 086 (2011).
[arXiv:1004.3736 [hep-th]].
%%CITATION = arXiv:1004.3736%%
}

%\MaldacenaRE
\lref\MaldacenaRE{
  J.~M.~Maldacena,
  ``The Large N limit of superconformal field theories and supergravity,''
Int.\ J.\ Theor.\ Phys.\  {\bf 38}, 1113 (1999), [Adv.\ Theor.\ Math.\ Phys.\  {\bf 2}, 231 (1998)].
[hep-th/9711200].
%%CITATION = HUTP-97-A097%%
}

%\WittenQJ
\lref\WittenQJ{
  E.~Witten,
  ``Anti-de Sitter space and holography,''
Adv.\ Theor.\ Math.\ Phys.\  {\bf 2}, 253 (1998).
[hep-th/9802150].
%%CITATION = hep-th/9802150%%
}

%\GubserBC
\lref\GubserBC{
  S.~S.~Gubser, I.~R.~Klebanov and A.~M.~Polyakov,
  ``Gauge theory correlators from noncritical string theory,''
Phys.\ Lett.\ B {\bf 428}, 105 (1998).
[hep-th/9802109].
%%CITATION = hep-th/9802109%%
}

%\GubserVV
\lref\GubserVV{
  S.~S.~Gubser and I.~R.~Klebanov,
  ``A Universal result on central charges in the presence of double trace deformations,''
Nucl.\ Phys.\ B {\bf 656}, 23 (2003).
[hep-th/0212138].
%%CITATION = hep-th/0212138%%
}

%\HofmanAR
\lref\HofmanAR{
  D.~M.~Hofman and J.~Maldacena,
  ``Conformal collider physics: Energy and charge correlations,''
JHEP {\bf 0805}, 012 (2008).
[arXiv:0803.1467 [hep-th]].
%%CITATION = arXiv:0803.1467%%
}

%\CordovaZEJ
\lref\CordovaZEJ{
  C.~Cordova, J.~Maldacena and G.~J.~Turiaci,
  ``Bounds on OPE Coefficients from Interference Effects in the Conformal Collider,''
JHEP {\bf 1711}, 032 (2017).
[arXiv:1710.03199 [hep-th]].
%%CITATION = arXiv:1710.03199%%
}

%\ChowdhuryVEL
\lref\ChowdhuryVEL{
  S.~D.~Chowdhury, J.~R.~David and S.~Prakash,
  ``Constraints on parity violating conformal field theories in $d=3$,''
JHEP {\bf 1711}, 171 (2017).
[arXiv:1707.03007 [hep-th]].
%%CITATION = arXiv:1707.03007%%
}

%\ZhiboedovOPA
\lref\ZhiboedovOPA{
  A.~Zhiboedov,
  ``On Conformal Field Theories With Extremal a/c Values,''
JHEP {\bf 1404}, 038 (2014).
[arXiv:1304.6075 [hep-th]].
%%CITATION = arXiv:1304.6075%%
}

%\SleightPCZ
\lref\SleightPCZ{
  C.~Sleight and M.~Taronna,
  ``Higher spin gauge theories and bulk locality: a no-go result,''
[arXiv:1704.07859 [hep-th]].
%%CITATION = arXiv:1704.07859%%
}

%\MaldacenaJN
\lref\MaldacenaJN{
  J.~Maldacena and A.~Zhiboedov,
  ``Constraining Conformal Field Theories with A Higher Spin Symmetry,''
J.\ Phys.\ A {\bf 46}, 214011 (2013).
[arXiv:1112.1016 [hep-th]].
%%CITATION = arXiv:1112.1016%%
}

%\AlbaYDA
\lref\AlbaYDA{
  V.~Alba and K.~Diab,
  ``Constraining conformal field theories with a higher spin symmetry in d=4,''
[arXiv:1307.8092 [hep-th]].
%%CITATION = arXiv:1307.8092%%
}

%\AlbaUPA
\lref\AlbaUPA{
  V.~Alba and K.~Diab,
  ``Constraining conformal field theories with a higher spin symmetry in $d > 3$ dimensions,''
JHEP {\bf 1603}, 044 (2016).
[arXiv:1510.02535 [hep-th]].
%%CITATION = arXiv:1510.02535%%
}

%\VenezianoYB
\lref\VenezianoYB{
  G.~Veneziano,
  ``Construction of a crossing - symmetric, Regge behaved amplitude for linearly rising trajectories,''
Nuovo Cim.\ A {\bf 57}, 190 (1968)..
}

%\MaldacenaIUA
\lref\MaldacenaIUA{
  J.~Maldacena, D.~Simmons-Duffin and A.~Zhiboedov,
  ``Looking for a bulk point,''
JHEP {\bf 1701}, 013 (2017).
[arXiv:1509.03612 [hep-th]].
%%CITATION = arXiv:1509.03612%%
}

%\GaryAE
\lref\GaryAE{
  M.~Gary, S.~B.~Giddings and J.~Penedones,
  ``Local bulk S-matrix elements and CFT singularities,''
Phys.\ Rev.\ D {\bf 80}, 085005 (2009).
[arXiv:0903.4437 [hep-th]].
%%CITATION = arXiv:0903.4437%%
}

%\ColemanAD
\lref\ColemanAD{
  S.~R.~Coleman and J.~Mandula,
  ``All Possible Symmetries of the S Matrix,''
Phys.\ Rev.\  {\bf 159}, 1251 (1967).
}

%\MaldacenaWAA
\lref\MaldacenaWAA{
  J.~Maldacena, S.~H.~Shenker and D.~Stanford,
  ``A bound on chaos,''
JHEP {\bf 1608}, 106 (2016).
[arXiv:1503.01409 [hep-th]].
%%CITATION = arXiv:1503.01409%%
}

%\SkvortsovLJA
\lref\SkvortsovLJA{
  E.~D.~Skvortsov and M.~Taronna,
  ``On Locality, Holography and Unfolding,''
JHEP {\bf 1511}, 044 (2015).
[arXiv:1508.04764 [hep-th]].
%%CITATION = arXiv:1508.04764%%
}

%\FitzpatrickYX
\lref\FitzpatrickYX{
  A.~L.~Fitzpatrick, J.~Kaplan, D.~Poland and D.~Simmons-Duffin,
  ``The Analytic Bootstrap and AdS Superhorizon Locality,''
JHEP {\bf 1312}, 004 (2013).
[arXiv:1212.3616 [hep-th]].
%%CITATION = arXiv:1212.3616%%
}

%\KomargodskiEK
\lref\KomargodskiEK{
  Z.~Komargodski and A.~Zhiboedov,
  ``Convexity and Liberation at Large Spin,''
JHEP {\bf 1311}, 140 (2013).
[arXiv:1212.4103 [hep-th]].
%%CITATION = arXiv:1212.4103%%
}

%\AldayEYA
\lref\AldayEYA{
  L.~F.~Alday, A.~Bissi and T.~Lukowski,
  ``Large spin systematics in CFT,''
JHEP {\bf 1511}, 101 (2015).
[arXiv:1502.07707 [hep-th]].
%%CITATION = arXiv:1502.07707%%
}

%\AldayOTA
\lref\AldayOTA{
  L.~F.~Alday and A.~Zhiboedov,
  ``Conformal Bootstrap With Slightly Broken Higher Spin Symmetry,''
JHEP {\bf 1606}, 091 (2016).
[arXiv:1506.04659 [hep-th]].
%%CITATION = arXiv:1506.04659%%
}

%\ChangKT
\lref\ChangKT{
  C.~M.~Chang, S.~Minwalla, T.~Sharma and X.~Yin,
  ``ABJ Triality: from Higher Spin Fields to Strings,''
J.\ Phys.\ A {\bf 46}, 214009 (2013).
[arXiv:1207.4485 [hep-th]].
%%CITATION = TIFR-TH-12-29%%
}

%\BoulangerOVA
\lref\BoulangerOVA{
  N.~Boulanger, P.~Kessel, E.~D.~Skvortsov and M.~Taronna,
  ``Higher spin interactions in four-dimensions: Vasiliev versus Fronsdal,''
J.\ Phys.\ A {\bf 49}, no. 9, 095402 (2016).
[arXiv:1508.04139 [hep-th]].
%%CITATION = arXiv:1508.04139%%
}

%\MeltzerRTF
\lref\MeltzerRTF{
  D.~Meltzer and E.~Perlmutter,
  ``Beyond $a=c$: Gravitational Couplings to Matter and the Stress Tensor OPE,''
[arXiv:1712.04861 [hep-th]].
%%CITATION = arXiv:1712.04861%%
}

%\AharonyDWX
\lref\AharonyDWX{
  O.~Aharony, L.~F.~Alday, A.~Bissi and E.~Perlmutter,
  ``Loops in AdS from Conformal Field Theory,''
JHEP {\bf 1707}, 036 (2017).
[arXiv:1612.03891 [hep-th]].
%%CITATION = arXiv:1612.03891%%
}

%\AldayNJK
\lref\AldayNJK{
  L.~F.~Alday,
  ``Large Spin Perturbation Theory for Conformal Field Theories,''
Phys.\ Rev.\ Lett.\  {\bf 119}, no. 11, 111601 (2017).
[arXiv:1611.01500 [hep-th]].
%%CITATION = arXiv:1611.01500%%
}

%\KravchukDZD
\lref\KravchukDZD{
  P.~Kravchuk,
  ``Casimir recursion relations for general conformal blocks,''
JHEP {\bf 1802}, 011 (2018).
[arXiv:1709.05347 [hep-th]].
%%CITATION = CALT-TH-2017-050%%
}

%\KarateevJGD
\lref\KarateevJGD{
  D.~Karateev, P.~Kravchuk and D.~Simmons-Duffin,
  ``Weight Shifting Operators and Conformal Blocks,''
[arXiv:1706.07813 [hep-th]].
%%CITATION = CALT-TH-2017-031%%
}

\Title{
\vbox{\baselineskip8pt
}}
{\vbox{
\centerline{Veneziano Amplitude of Vasiliev Theory}
\vskip.1in
}}

\vskip.1in
\bigskip
\centerline{Gustavo J. Turiaci$^1$ and Alexander Zhiboedov$^2$}
\vskip.2in 
\centerline{\it $^1$ Department of Physics, Princeton University, Princeton, NJ 08544, USA}
\centerline{\it $^2$
Department of Physics, Harvard University, Cambridge, MA 02138, USA}

\vskip.7in \centerline{\bf Abstract}{ 
\vskip.2in
We compute the four-point function of scalar operators in CFTs with weakly broken higher spin symmetry at arbitrary 't Hooft coupling. We use the known three-point functions in these theories, the Lorentzian OPE inversion formula and crossing to fix the result up to the addition of three functions of the cross ratios. These are given by contact Witten diagrams in AdS and manifest non-analyticity of the OPE data in spin. We use Schwinger-Dyson equations to show that such terms are absent in the large $N$ Chern-Simons matter theories. The result is that the OPE data is analytic in spin up to $J=0$. 
}
 
\Date{February 2018}

%\listtoc\writetoc
\vskip 1.5in \noindent

%\vskip 3.5in 

\newsec{Introduction}

The AdS/CFT correspondence \refs{\MaldacenaRE\WittenQJ-\GubserBC} relates four-dimensional higher spin gravity in $AdS_4$ \refs{\VasilievEN\VasilievBA-\VasilievEV} to a three-dimensional conformal field theory with almost conserved higher spin currents \refs{\KlebanovJA\SezginRT\GiombiWH\GiombiVG\BoulangerOVA-\SleightDBA}. 
These are large $N$ CFTs with the anomalous dimensions of higher spin currents being suppressed by ${1 \over N}$. A famous example of such theories is given by Chern-Simons gauge fields coupled to the fundamental matter \refs{\GiombiKC\AharonyNS \AharonyNH\AharonyJZ\GurAriIS-\Minwalla}. The presence of slightly broken higher spin symmetry makes the theory solvable in the planar limit. In spite of this fact deriving consequences of this broken symmetry has remained elusive.\foot{In contrast, when the higher spin symmetry is unbroken the correlators of the theory are completely fixed \refs{\MaldacenaJN\AlbaYDA\AlbaUPA-\DidenkoTV}.} In particular, even in the planar limit only the three-point functions are known in this class of theories \refs{\AharonyJZ\GurAriIS-\MaldacenaSF}. Existing methods for studying these theories, be it higher spin Ward identities \MaldacenaSF\ or Schwinger-Dyson equations \GiombiKC, become very complicated at the level of the four-point functions. Recent discussions of the four-point functions in the context of the higher spin gauge/CFT duality include \refs{\BekaertCEA\BekaertTVA\BekaertEZC\SkvortsovLJA\SleightPCZ-\SleightCAX}.\foot{Correlators of the Vasiliev theory duals have a stringy feature that exchange of higher spin currents in the $s$-channel is equal to the exchange of higher spin currents in the $t$-channel, see, e.g., \AldayOTA. The same is true for the usual Veneziano amplitude \VenezianoYB. It would be interesting to understand if there is a worldsheet formulation of the Vasiliev theory which makes this fact manifest, see, e.g., \ChangKT.} In particular, in \PRAKASH\ the Schwinger-Dyson equations were used to compute the four-point correlation functions in the double soft collinear limit.\foot{We thank Simone Giombi for bringing our attention to this paper.} 

In this paper we compute scalar four-point functions in large $N$ CFTs with slightly broken higher spin symmetry at arbitrary 't Hooft coupling. We use the known results for the three-point functions \MaldacenaSF\ together with analyticity in spin argument \Simon\ and crossing to fix the answer up to three arbitrary functions of the 't Hooft coupling. We then use the Schwinger-Dyson equations \Minwalla\ to fix these constants to zero. 

The result of our analysis is the following. Recall that the theories of interest come in two guises, the so-called {\it quasi-fermion} theory and {\it quasi-boson} theory. Both theories are characterized by the two-point function of stress tensor $\tilde N$ and by the 't Hooft coupling $\tilde \lambda$. They have the same spectrum of higher spin currents but different spectra of scalar operators. In the quasi-fermion theory there is an operator of dimension $\Delta_{qf} = 2 +O(1 / \tilde N)$, whereas in the quasi-boson theory an operator of dimension $\Delta_{qb} = 1+O(1 / \tilde N)$.  Based on general arguments we find a three-parametric family of correlators. We then fix these parameters using the particular realization of Chern-Simons matter theory, for which they vanish. We do not know if there are other examples where these parameters take non-zero values. 

In the quasi-fermion theory the result takes the following form
\eqn\resultfourQF{\eqalign{
\la O_{qf}(x_1)O_{qf}(x_2)O_{qf}(x_3)O_{qf}(x_4)\ra_{conn} &= {1 \over \tilde N} {f_{ff}(u,v)\over x_{13}^4 x_{24}^4} ,
}}
where we wrote only the connected piece and $f_{ff}(u,v)$ stands for the four-point function in the free fermion theory. In writing \resultfourQF\ we normalized the two-point function to $\la O O\ra = {1 \over x^{2 \Delta}}$. Remarkably the four-point function of scalars in the quasi-fermion theory is simply given by the free fermion result and does not depend on $\tilde \lambda_{qf}$. By taking $\tilde \lambda_{qf} \to \infty$ limit we obtain the prediction that the four-point function in the critical $O(N)$ model is proportional to the free fermion one. This was indeed found to be the case in \LeonhardtDU\ by the direct evaluation of Feynman diagrams. 

In the quasi-boson theory the scalar four-point function takes the following form
\eqn\resultfourQB{\eqalign{
&\la O_{qb}(x_1)O_{qb}(x_2)O_{qb}(x_3)O_{qb}(x_4)\ra_{conn} = {1 \over \tilde N} {f_{qb}(u,v)\over x_{13}^2 x_{24}^2} ,\cr
f_{qb}(u,v) &=  f_{fb}(u,v) - {\tilde \lambda^2_{qb} \over 1 + \tilde \lambda^2_{qb} } {8 \over \pi^{5/2}}\left( \bar D_{1,1,{1 \over 2}, {1 \over 2}}(u,v) + \bar D_{1,1,{1 \over 2}, {1 \over 2}}(v,u) + {1 \over u} \bar D_{1,1,{1 \over 2}, {1 \over 2}}({1 \over u},{v \over u}) \right) ,
}}
where $f_{fb}(u,v)$ stands for the correlator in the free boson theory. The second term is a sum over exchange diagrams in AdS$_4$ for the $\phi^3$ vertex. By taking $\tilde \lambda_{qb} \to \infty$ limit we obtain a prediction from \resultfourQB\ for the four-point function in the critical Gross-Neveu theory, which to our knowledge is new.

The rest of the paper is organized as follows. In section 2 we present some general CFT arguments that relate the full correlator and its double discontinuity. In section 3 we use these arguments to fix the correlators in large $N$ CFTs with weakly broken higher spin symmetry up to three unknown functions of the 't Hooft coupling. In section 4 we use the Schwinger-Dyson equations in the Chern-Simons matter theories to fix these coefficients to zero. We finish with some open directions and conclusions. We collect some technical details in the appendices.

\newsec{From Double Discontinuity to Full Correlator}

We will find it useful to use a couple of simple theorems that could be proven for generic CFTs that relate the data that is contained in the double discontinuity of the correlator to the full correlator. 

We consider a four-point function of identical primary operators. As was widely discussed recently it is useful to expand the correlator in terms of the complete basis of conformal partial waves $F_{\Delta, J}(x_i)$ labeled by spin $J$ and dimension ${d \over 2} + i \nu$
\eqn\expansion{\eqalign{
G = \la O(x_1) O(x_2) O(x_3) O(x_4) \ra = ( {\rm non - norm}.) + \sum_{J=0}^{\infty} \int_{{d \over 2} - i \infty}^{{d \over 2} + i \infty} {d \Delta \over 2 \pi i} c_{\Delta, J} F_{\Delta, J}(x_i) \ ,
}}
where the non-normalizable part includes scalar operators with dimension $\Delta' < {d \over 2}$ that appear in the OPE of $O(x_1) O(x_2)$ (see appendix B.2 of \SimmonsDuffinNUB). By deforming the contour of integration we recover the usual OPE expansion. 

The coefficients of this expansion $c_{\Delta, J}$ could be written in terms of the kernel integrated against the  double discontinuities of the correlator $\la[O(x_2) , O(x_3)] [O(x_1) , O(x_4)] \ra$ and $\la [O(x_2), O(x_4)] [O(x_1) , O(x_3)]\ra$ as discovered by Caron-Huot \Simon\ (see also \refs{\SimmonsDuffinNUB})\foot{This result could be alternatively seen in the large spin expansion \refs{\AldayVKK\AldayNJK-\AharonyDWX}.}. Schematically, the relation takes the following form
\eqn\coefficient{
c_{\Delta, J} = \int_0^1 d z d \bar z \ K^t_{\Delta, J}(z, \bar z) \la[O_2 , O_3] [O_1 , O_4] \ra  + (-1)^J \int_{-\infty}^{0} d z d \bar z \ K^u_{\Delta, J}(z, \bar z) \la[O_2 , O_4] [O_1 , O_3] \ra ,
}
where $K^t_{\Delta, J}(z, \bar z) $ and $K^u_{\Delta, J}(z, \bar z) $ are explicitly known. This formula follows from boundedness of the correlator in the Regge limit. At finite $N$ this formula works for $J>1$ \Simon. In the large $N$ limit it works for $J>2$. The reason for this difference is that the bound on the Regge limit is weaker in the large $N$ limit \MaldacenaWAA.

What we need from this formula is the fact that given double discontinuity of the correlator we could in principle find the OPE data of the correspondent spinning operators appearing in the OPE of external operators. We can prove the following simple but useful statements.

\vskip 0.1in \noindent

{\bf Theorem 1:} In a CFT, the four-point correlator $G$ of identical scalars is completely fixed by its double discontinuity.

\vskip 0.1in \noindent

\noindent Imagine that it is not the case. It means that we could construct two solutions to crossing $G_1$ and $G_2$ such that they have 
\eqn\doubledisc{
{\rm dDisc}[G_1] = {\rm dDisc}[G_2] 
}
in every channel.\foot{As with the usual OPE there are three different ways to construct the double commutator.} Using the inversion formula it means that in the OPE expansion of $G_1$ and $G_2$ in every channel all the operators with $J>1$ appear with the same three-point coefficients.\foot{Operators with $J=1$ are absent in the OPE of identical operators.} Let us consider the difference between the two correlators
\eqn\differenceA{
\delta G = G_1 - G_2 .
}
By construction this difference is crossing symmetric and bounded in the Regge limit. Moreover, it admits a convergent OPE expansion with only scalar primary operators appearing in every channel. We therefore get the crossing equation for $\delta G(u,v)$
\eqn\crossingequationA{
{1 \over u^{\Delta}} \sum_{\Delta'} c_{\Delta'} g_{\Delta', 0}(u,v) = {1 \over v^{\Delta}} \sum_{\Delta'} c_{\Delta'} g_{\Delta', 0}(v,u).
}
There are two differences between \crossingequationA\ and the usual crossing equations. First, the only operators that appear in the expansion are scalars. Second, $c_{\Delta}$ are not necessarily positive since we are considering a difference between two correlators \differenceA. It is very easy to show that such solutions to crossing do not exist.

Note that $\delta G$ by construction has zero double discontinuity in every channel. It means \Simon\ that it contains only scalar operators with dimensions $\Delta' = 2 \Delta + 2 n$ in every channel, which is equivalent to the statement that it is Casimir biregular in this case. To see this recall that a given operator contributes to the double discontinuity with the factor $\sin^2 ({\pi (\Delta' - 2 \Delta - J) \over 2})$. Requiring vanishing of this pre-factor implies that $\Delta' =2 \Delta + J + 2 n $. Therefore, the crossing equation takes the form\foot{Another way to see it is to consider the limit $u,v \to 0$ of \crossingequationA . By the standard arguments 
\refs{\FitzpatrickYX\KomargodskiEK-\AldayEYA} all the Casimir-singular terms (terms which become singular after the action of the Casimir operator, see, e.g., \SimmonsDuffinWLQ) on one side should be reproduced by higher spin operators on the other side. Since we have only scalar operators the conclusion is that $\delta G$ is Casimir regular in both channels, or biregular in the terminology of \SimmonsDuffinWLQ.}
\eqn\crossingequationB{
{1 \over u^{\Delta}} \sum_{n} c_{n} g_{2 \Delta + 2n, 0}(u,v) = {1 \over v^{\Delta}} \sum_{n} c_{n} g_{2\Delta + 2n, 0}(v,u) \ .
}

However, this is precisely a type of problem considered by Heemskerk, Polchinski, Penedones and Sully in \HPPS\ who analyzed solutions to the large $N$ crossing. The essential idea of \HPPS\ was to match the discontinuities of the equation \crossingequationB\ around $u = 0$ (and/or $v = 0$). 

Indeed consider, for example, a discontinuity along $v = 0$, namely $\delta G|_{v e^{i \pi}} - \delta G|_{v e^{-i \pi}}$ for small $u$. The RHS of \crossingequationB\ has an expansion in integer powers of $v$ only and, thus, produces $0$. In the LHS on the other hand each conformal block has a singularity $\log v$ which produces non-trivial discontinuity $2 \pi i$. Requiring that it is zero for every $u$ sets $c_n = 0$. Indeed, different primaries come with different powers of $u^n + ...$ and, therefore, we have to set all the coefficients to $0$. Therefore, $\delta G = 0$ and
\eqn\conclusionA{
G_1 = G_2  \ , 
}
which is what we wanted to show.

In \HPPS\ this argument fails because $\log v$ terms are present in the RHS as well, due to the anomalous dimensions of double trace operators when we expand the correlator in ${1 \over N}$. These terms will be important below. In the argument above such terms are absent due to the condition ${\rm dDisc}[\delta G] = 0$.

In our paper we will be dealing with the large $N$ theories and therefore a slightly modified version of the statement above would be useful for us.

\vskip 0.1in \noindent

{\bf Theorem 2:} In a large $N$ CFT, the planar four-point correlator $G$ of identical scalars is fixed by its double discontinuity up to a function which is a sum of three contact AdS Witten diagrams with arbitrary coefficients. 

\vskip 0.1in \noindent

Again, imagine that it is not the case.  We consider two solutions to the large $N$ crossing $G_1$ and $G_2$ with the same discontinuity in all channels and consider their difference $\delta G$. Again it admits the large $N$ OPE expansion, this time, however with operators of $J=0$ and $J=2$ potentially appearing in every channel. 

Since by construction the double discontinuity of $\delta G$ is zero the only operators that could appear in the OPE of $\delta G$ are the usual double trace operators of $J=0$ and $J=2$. This is, however, precisely the problem solved by \HPPS. They showed that purely double trace solutions to the large $N$ crossing that are bounded in spin are in one-to-one correspondence with the contact interactions in AdS. In particular, in our case the relevant vertices in AdS are $\phi^4$, $(\pa \phi)^4$ and $\phi^2 (\pa^3 \phi)^2$. We conclude that 
\eqn\conclusion{
G_1 = G_2 + c_1 G_{\phi^4}^{AdS} + c_2 G_{(\pa \phi)^4}^{AdS} + c_3 G_{\phi^2 (\pa^3 \phi)^2}^{AdS}
}
\noindent which was to be demonstrated. 

Let us reiterate the differences between the two arguments that allow for subtraction terms in the large $N$ case only. One obvious difference, to be emphasized again, is that the bound on the large $N$ Regge limit leaves some freedom in the $J=2$ sector. But what about $J=0$ and $G_{\phi^4}^{AdS}$? The reason is that $G_{\phi^4}^{AdS}$ is a solution to crossing only to leading order in ${1 \over N}$ and would not be a solution to a finite $N$ crossing equation \crossingequationA\ which we used in the previous argument for a generic CFT. 

We will use  \conclusion\ to bootstrap some correlators in the large $N$ Chern-Simons matter theory below.

\newsec{Conformal Field Theories With Weakly Broken Higher Spin Symmetry}

We would like to apply the arguments of the previous section to a class of particularly simple CFTs. We consider large $\tilde N$ CFTs\foot{We use $\tilde N$ to denote the two-point function of stress tensors in an abstract CFT. By $N$ we denote the number of colors in the Chern-Simons matter theories.} with the following spectrum of single trace operators. First, we have a set of almost conserved higher spin currents $j_s$ with dimensions
\eqn\dimalmcons{
\Delta_s = s+1 + O({1 \over \tilde N}) \ , ~~~ s>0 \ .
}
These include the stress tensor for $s=2$. Second, we have a scalar operator which distinguishes between the two type of theories that we consider. In the so-called {\it quasi-fermion} theory it has the dimension 
\eqn\quasif{
\Delta_{qf} = 2  + O({1 \over \tilde N}) \ .
}
In the {\it quasi-boson} theory it has the dimension
\eqn\quasib{
\Delta_{qb} = 1  + O({1 \over \tilde N}) \ .
}
Each theory is as well characterized by the correspondent 't Hooft coupling $\tilde \lambda_{qf}$ and $\tilde \lambda_{qb}$ which could take arbitrary values. When $\tilde \lambda_{qf} = 0$ we have the free fermion CFT, and $\tilde \lambda_{qb} = 0$ we have the theory of free bosons. Microscopically, the quasi-fermion (quasi-boson) theory could be realized as the Chern-Simons theory coupled to fermions (bosons) in the fundamental representation of the gauge group  \refs{\GiombiKC\AharonyNS \AharonyNH\AharonyJZ\GurAriIS-\Minwalla}. 

In this paper we are primarily interested in computing the four-point functions of the scalar operators \quasif, \quasib\ in the planar limit and at arbitrary 't Hooft coupling 

\eqn\fourpointcorr{\eqalign{
\la O_{qf}(x_1)O_{qf}(x_2)O_{qf}(x_3)O_{qf}(x_4)\ra &= {\rm disc} + {1 \over \tilde N} {f_{qf}(u,v)\over x_{13}^4 x_{24}^4} , \cr
\la O_{qb}(x_1)O_{qb}(x_2)O_{qb}(x_3)O_{qb}(x_4)\ra &= {\rm disc} + {1 \over \tilde N} {f_{qb}(u,v)\over x_{13}^2 x_{24}^2} \ ,
}}
where ${\rm disc} = {1 \over x_{12}^{2 \Delta} x_{34}^{2 \Delta}} +  {1 \over x_{13}^{2 \Delta} x_{24}^{2 \Delta}} +  {1 \over x_{14}^{2 \Delta} x_{23}^{2 \Delta}} $ is the disconnected piece. Note also that we normalized the two-point functions as $\la O(x) O(0)\ra =  x^{-2 \Delta}$. Parameter $\tilde N$ parametrizes the two-point function of stress tensors in the theory, see for example \GurAriIS.  Our goal, thus, is to compute the non-trivial functions $f_{qf}(u,v)$ and $f_{qb}(u,v)$.

On general grounds the correlators \fourpointcorr\ are crossing symmetric and admit an OPE expansion. Since we are dealing with the large $\tilde N$ CFT the operators that appear in the OPE to leading order in ${1 \over \tilde N}$ are single trace operators and double trace operators. An important simplification for our task is that the contribution of single trace operators is known since the correspondent three-point couplings were computed before \MaldacenaSF. Let us quickly review these results.

\subsec{Correlator in the Quasi-fermion Theory}

In this case the relevant three-point functions are proportional to the ones in the theory of the free fermion \MaldacenaSF. More precisely, the normalization independent square of the three-point function\foot{It is given by the ratio $ {\la O_{qf} O_{qf} j_s\ra^2 \over \la O_{qf} O_{qf}\ra^2 \la j_s j_s \ra}$ which is invariant under rescaling of operators involved.} takes the form
\eqn\qfthreepoint{
c^2_{s, qf} ={1 \over \tilde N} c^2_{s,ff} , ~~~ s \geq 0 \ ,
}
which is to say that the three-point functions are the same as in the theory of $\tilde N$ free fermions. In the theory of Chern-Simons plus fermions $\tilde N$ is a known function of the rank of the gauge group and the 't Hooft coupling $\lambda$ as we review below. In particular, the three-point function of scalars $\la O_{qf} O_{qf} O_{qf} \ra = 0$ vanishes in the quasi-fermion theory at separated point.

We would like to construct a solution of crossing $\tilde f_{qf}(u,v)$ such that 
\eqn\agreementQF{
{\rm dDisc}[f_{qf}(u,v)] = {\rm dDisc}[\tilde f_{qf}(u,v)] 
}
in all channels. This task is completely trivial in the quasi-fermion theory. Note that all the three-point functions are proportional to the ones in the free fermion theory \qfthreepoint. The correct function $\tilde f_{qf}(u,v)$ that satisfies \agreementQF\
is proportional to the free fermion answer $f_{ff}(u,v)$ 
\eqn\freefermion{\eqalign{
f_{ff}(u,v) &= {1 + u^{5/2} + v^{5/2} - u^{3/2}(1+v) - v^{3/2}(1+u)-u -v \over u^{3/2} v^{3/2}} \ .
}}

By the argument above \conclusion\ we conclude that 
\eqn\identityQF{
f_{qf}(u,v) = f_{ff}(u,v) + c_1^{qf} f_{\phi^4}^{AdS}(u,v) + c_2^{qf} f_{(\pa \phi)^4}^{AdS}(u,v) + c_3^{qf} f_{\phi^2 (\pa^3 \phi)^2}^{AdS}(u,v).
}

In this way we conclude that the four-point function of scalars in the quasi-fermion theory is 
fixed up to three unknown functions of the 't Hooft coupling $c_i$. It could well be that these are not consistent with weakly broken higher spin symmetry but this involves analysis of complicated higher spin Ward identities (see appendix B), which we leave for the future work.

Instead we will use Schwinger-Dyson equations to fix these unknown coefficients in the large $N$ Chern-Simons matter theories which are concrete realizations of CFTs with slightly broken higher spin symmetry.

\subsec{Correlator in the Quasi-boson Theory}

The situation in the quasi-boson theory is slightly more complicated. For spinning operators we have normalization independent three-point functions
\eqn\qbthreepoint{
c^2_{s, qb} = {1 \over \tilde N} c^2_{s,fb} , ~~~ s > 0 ,
}
where again $c^2_{s,fb}$ refers to the result in a theory of a single free boson. The three-point function of scalars, however, obeys the following relation
\eqn\qbthreepointS{
c^2_{0, qb} = {1 \over \tilde N} {1 \over 1 + \tilde \lambda_{qb}^2} c^2_{0,fb} .
}
Note the presence of an extra factor $ {1 \over 1 + \tilde \lambda_{qb}^2}$ which depends on the 't Hooft coupling in the quasi-boson theory.

Again we would like to construct a function that satisfies the crossing equations, bound on the Regge limit and has the correct double discontinuity. In this case a natural starting point would be to take the free boson answer
\eqn\freefermion{
 f_{fb}(u,v) = 4 {1 + u^{1/2} + v^{1/2} \over u^{1/2} v^{1/2}} \ .
}
This, however, does not have the double discontinuity of the quasi-boson due to the mismatch of the contribution of the scalar operator \qbthreepointS. To correct this we can add a sum of $\phi^3$ exchange diagrams in AdS with the correct coefficient so that the scalar operator contributes with the correct coefficient. Exchange diagrams in AdS have the property that the only single trace operator that appears in their OPE expansion has the quantum numbers of the exchanged particle. In this way we can correct the coefficient in front the scalar operator without affecting operators with non-zero spin. Moreover, exchange diagrams in AdS have nice behavior in the Regge limit and therefore this correction satisfied the bound on Regge as well.  Using the results of \PenedonesUE\ for the arbitrary exchanged diagram,\foot{The relevant result is given by formulas (40-41) in that paper. Notice that it is singular in the limit, when $\Delta_i = \Delta = 1$ and $d=3$. This singularity, however, is just a trivial overall factor which we can simply drop. In the context of Vasiliev theory this was discussed, for example, in \SezginPT.} the result takes the following form\foot{One way to define a CFT with slightly broken higher spin symmetry is through the non-conservation of a higher spin current. It takes a schematic form $\pa_{\mu} j^{\mu}_s = a_2 j j + a_3 j j j$, see \MaldacenaSF\ for details. In our considerations we implicitly set the triple trace term $a_3=0$. The effect of the triple trace deformation $a_3$ is to change the coefficient in front of the sum of $\bar D$-functions.} 
\eqn\quasiboson{\eqalign{
f_{qb}(u,v) &= f_{fb}(u,v) - {\tilde \lambda^2_{qb} \over 1 + \tilde \lambda^2_{qb} } {8 \over \pi^{5/2}}\left( \bar D_{1,1,{1 \over 2}, {1 \over 2}}(u,v) + \bar D_{1,1,{1 \over 2}, {1 \over 2}}(v,u) + {1 \over u} \bar D_{1,1,{1 \over 2}, {1 \over 2}}({1 \over u},{v \over u}) \right)  \cr
&+ c_1^{qb} f_{\phi^4}^{AdS}(u,v) + c_2^{qb} f_{(\pa \phi)^4}^{AdS}(u,v) + c_3^{qb} f_{\phi^2 (\pa^3 \phi)^2}^{AdS}(u,v) .
}}
One can check that the expression above is crossing symmetric. The relevant properties of the $\bar D$-functions could be found for example in appendix C of \DolanIY. The coefficient in front of the $\bar D$-functions is such that the three-point function \qbthreepointS\ is correctly reproduced.

To conclude, both in the case of the quasi-fermion and quasi-boson theory the correlation function of the four scalar operators is fixed up to three unknown functions of 't Hooft coupling. In the next section we turn to the particular realization of these theories, namely the Chern-Simons matter theories. There we will show that $c_i  = 0$ both for the quasi-fermion and the quasi-boson theories.

\newsec{Schwinger-Dyson Approach to Chern-Simons Vector Models}
In this section we study a specific realization of theories to which the analysis above applies, namely Chern-Simons gauge fields coupled to either a fundamental fermion or a fundamental boson in the planar limit. The main result of this section can be summarized as 
\eqn\cs{
c_i^{qf/qb}=0 \ ,
}
with these parameters defined as in \identityQF~and \quasiboson. Instead of using a CFT-based approach, we use here the diagrammatics of the theories, writing a solution to the Schwinger-Dyson equations that allows us to compute the scalar four-point function for arbitrary 't Hooft coupling and explicitly verify \cs~together with \identityQF~and \quasiboson. 

\subsec{Setup} 

First, we consider the quasi-fermion theory, namely the $U(N)_k$ Chern-Simons field of level $k$ coupled to a fundamental fermion $\psi^i$, $i=1,\ldots, N$. The explicit Lagrangian and conventions can be found in \GiombiKC. The goal of this section is to compute the scalar four-point function for this specific theory in the limit of large $N$ and arbitrary 't Hooft coupling $\lambda=N/k$. We will be interested in the operator $O_{qf} \sim \bar{\psi} \psi$ and the connected part of the following correlator
\eqn\fourptmomentum{
\la O_{qf}(q_1)O_{qf}(q_2)O_{qf}(q_3)O_{qf}(q_4)\ra ,
}
where the operators are inserted with definite momenta $q_i$.

Since we resum the Feynman diagrams of the theory, the calculations in this section are best done in momentum space. For technical reasons that become clear below we focus on a particular collinear kinematic regime in which all four external operators have momentum aligned along the same component $q_i^\pm =0$ and $q_i^3\neq 0$ with $i=1,\ldots, 4$. To ease the notation momenta along the direction $3$ will be denoted simply by $q$. 

The analysis in the previous section was done in the normalization $\langle O_{qf} (x) O_{qf}(0) \rangle=x^{-2\Delta}$. Using the results of \GurAriIS\ the correct definition of the scalar operators is
\eqn\normO{
O_{qf} \equiv   {\pi^{3/2} \lambda^{1/2} \over 2 N^{1/2} \tan^{1/2}{\pi \lambda \over 2}}   : \bar{\psi}^i \psi_i :. 
}
Another ingredient of our proposal is the free fermion four-point function which we need in momentum space in a normalization consistent with \freefermion. This can be computed explicitly and the result is
\eqn\ffinfs{
F_{ff}(q_1,q_2,q_3,q_4)={\pi^4 \over 4}{ q_1 |q_1|+q_2|q_2|+q_3|q_3|+q_4 |q_4| \over (q_1+q_2)(q_1+q_3)(q_2+q_3)} \delta(\sum_i q_i).
} 
Recalling \identityQF, in momentum space we will verify the following 
\eqn\prop{
\la O_{qf}(q_1)O_{qf}(q_2)O_{qf}(q_3)O_{qf}(q_4)\ra_{\rm conn.} = {1 \over \tilde{N}}~ F_{ff}(q_1,q_2,q_3,q_4) \ ,
}
where the relation between the parameters $\tilde{N}$ and $\tilde{\lambda}$, appearing in the general approach based on softly-broken higher spin symmetry, and the parameters $N$ and $\lambda$ of the theory were found in \GurAriIS~to be 
\eqn\param{
\tilde{N} = 2 N { \sin \pi \lambda \over \pi \lambda},~~~\tilde{\lambda} = \tan {\pi \lambda \over 2} \ .
}

The contact diagrams appearing in \identityQF\ can also be written in momentum space. The expressions will not be necessary and therefore we will not go into detail but the outline of the calculation is simple. In AdS$_4$ the bulk fields dual to $\Delta=1$ or $\Delta=2$ are conformally coupled. Therefore one can perform the calculation in flat space and account for the scale transformation. This calculation is nicely explained in \ArkaniHamedBZA~for the case of dS$_4$. Moreover since $\Delta=1$ and $\Delta=2$ correspond to standard and alternate quantization of the bulk field, each contact diagram for $\Delta=1$ is given by the Legendre transform of the respective one for $\Delta=2$ \GubserVV.

In order to check \prop\ we follow \refs{\GiombiKC\AharonyNH-\GurAriIS} conventions and work in the light-cone gauge $A_-=0$. The simplicity of the diagrams contributing in the 't Hooft limit and the fact that the gauge boson propagator does not get corrected, allow us to write Schwinger-Dyson equations that resum all diagrams we need. In the case of collinear momenta \refs{\Minwalla,\PRAKASH}, this is enough to compute \fourptmomentum\ for arbitrary 't Hooft coupling.

We will follow the method used in \PRAKASH. To compute \fourptmomentum\ we need some partial results that were already obtained in the literature. We give here a summary of the necessary ingredients and leave detailed definitions and explicit formulas to appendix A. 

First of all we need the exact fermion propagator in the 't Hooft limit $S(p)$ for arbitrary $\lambda$. This was computed in \GiombiKC~and found to be two-loop exact, see (A.2). With $S(p)$ one can write Schwinger-Dyson equation that resums all diagrams contributing to the form factor $\langle O_{qf}(q) \psi_i(k) \bar{\psi^j}(-p)\rangle$.\foot{ In the planar limit this is enough since multiparticle contributions are suppressed by $N$.} This was solved in \GurAriIS~and following their conventions we will denote this vertex by $V(q,p)$, which we define in (A.3) using our convention for the normalization of $O_{qf}$. 

\ifig\SDfig{Diagrammatic representation of the Schwinger-Dyson equation defining the kernel $\Gamma$, denoted as a blue blob, that resums ladder diagrams. Lines denote exact fermion propagators and wavy line gluon ones. The external momenta $k$ and $r$ are arbitrary while $q$ points in the 3-direction.} {\epsfxsize 4.5in\epsfbox{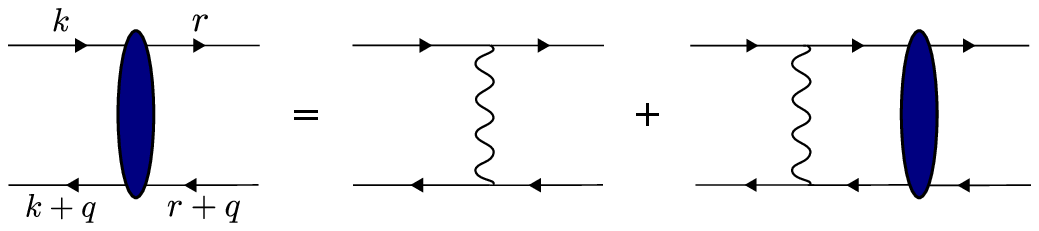}}

To compute the four-point correlator of $O_{qf}$ we further need to compute the four-fermion vertex. Therefore, we need to solve the Schwinger-Dyson equation that resums all diagrams involving four dressed fermionic external legs. Following \refs{\GiombiKC,\Minwalla}, we can define this object by an effective action which at tree level would reproduce the non-perturbative four-point function of the fermions for arbitrary $\lambda$
\eqn\Gammadef{
S_{\rm eff} = -{1 \over 2}\int { d^3k d^3q d^3r \over  (2\pi)^9} \Gamma_{AB}(k,q,r) \bar{\psi}^i (-k-q) \gamma^A \psi_j (r+q) \bar{\psi}^j(-r) \gamma^B \psi_i (k).
} 
This corresponds to the blue blob in \SDfig. As defined in \Minwalla~the object $\Gamma$ has four spinor indices but we follow the convention of \PRAKASH~of using the compact notation $\Gamma_{AB}(k,q,r) \gamma^A \otimes \gamma^B$, where $A=1,\ldots, 4$ and $\gamma^4={\bf 1}$. 

In the light-cone gauge we only need to consider ladder diagrams made up of gluon exchanges, as long as we use the dressed fermion propagators. This implies that the Schwinger-Dyson equation for the kernel $\Gamma_{AB}(k,q,r)$ can be written 
\eqn\SDeq{\eqalign{
\Gamma_{AB}(k,q,r) \gamma^A \otimes \gamma^B &= {2 \pi \lambda i \over N} { \gamma^{[+|}\otimes \gamma^{|3]} \over r^+ - k^+}\cr
&+ \int {d^3 \ell \over (2\pi)^3} {2 \pi i \lambda \over k^+-\ell^+ } \Gamma_{CD}(\ell,q,r)  \gamma^{[+|} S(\ell+q) \gamma^{C} \otimes \gamma^{D} S(\ell) \gamma^{|3]},
}}
which is represented by the diagram shown in \SDfig. The brackets in the gamma matrices denote antisymmetrization. This equation was derived in \GiombiKC~, the solution was found in \Minwalla, and was used to compute the scalar four-point function in \PRAKASH. We give the explicit expression for it in appendix A, equations (A.4) to (A.9). One can see that $\Gamma_{AB} \sim O(1/N)$, and therefore each term in \SDeq~contributes equally. Our normalization for $\Gamma_{AB}$ in (A.4) differs from \PRAKASH\ but we have checked that (A.4)-(A.9) do solve the Schwinger-Dyson equation \SDeq. 

\ifig\ABfig{Diagrams contributing to the four-point function of $O_{qf}(q)$. The momenta of the external particles $q$ is along direction $3$ and always incoming. Diagrams B include the four-fermion vertex which is denoted by a blue blob. We omit the other 5 permutations of diagram A and show only nonequivalent configurations of diagram B.}{\epsfxsize4in\epsfbox{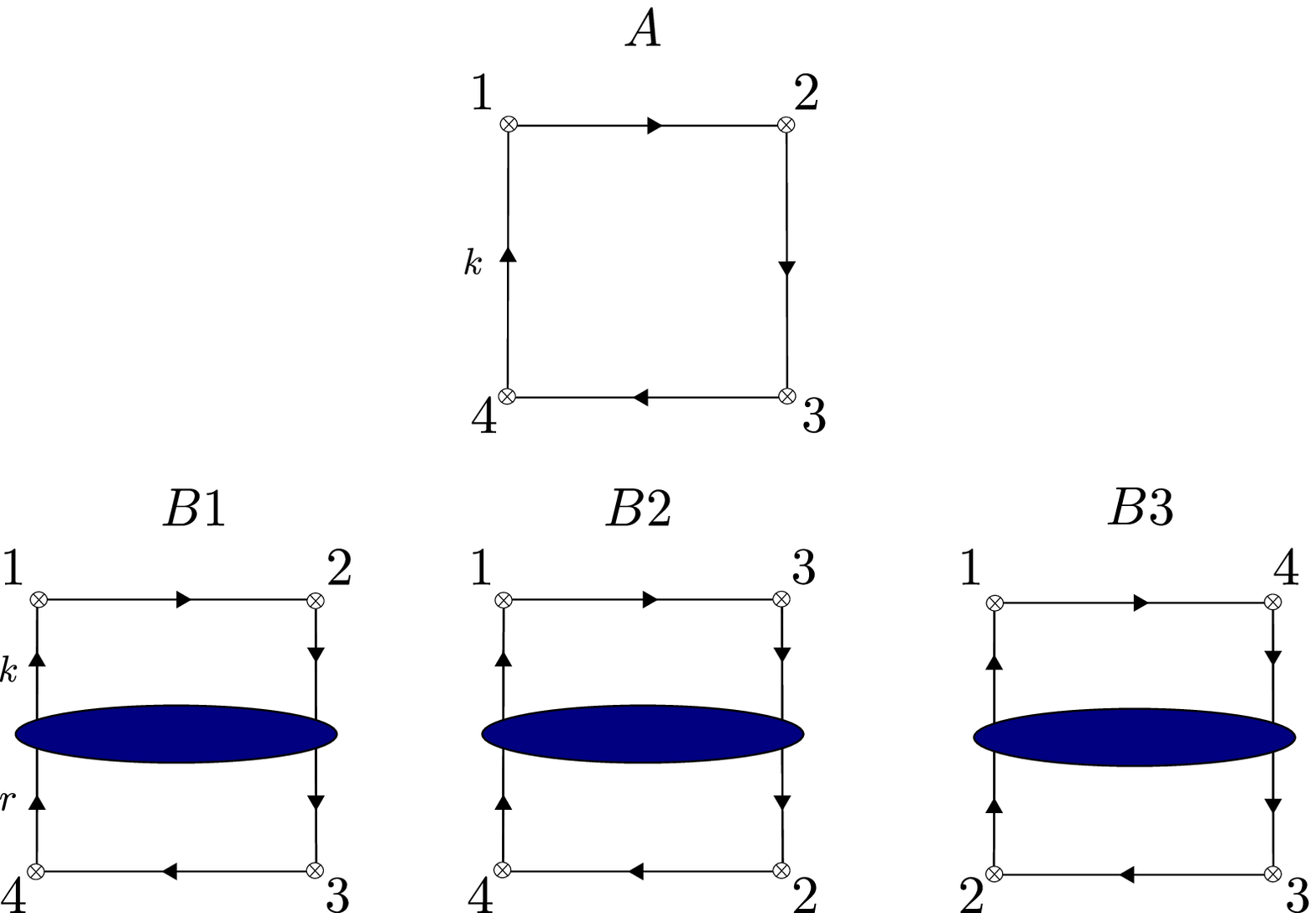}}

Now we can enumerate the diagrams appearing in the four-point function, which are shown in \ABfig. Diagrams A are ones in which the four-fermion vertex $\Gamma$ is not inserted. We denote the $O_{qf}$ form factor by the symbol ``$\otimes$" following \GurAriIS\ and we take the momentum $q_i$, $i=1,\ldots, 4$ to be ingoing and collinear for all external insertions. In the figure we indicated a choice of ordering of the external momenta and one should sum over six permutations of the momenta. These diagrams involve a single integral over momentum that we denote $k$ and a single color loop making them of order $1/N$. In (A.10) we show an example of an integrand for the permutation shown in the picture. 

In the second line of the figure we show diagrams, labeled as B, for which there is a single insertion of the $\Gamma$ vertex. In these diagrams there are two integrals over arbitrary momenta that we label $k$ and $r$. There are also two color loops giving an $N$ enhancement with respect to A but the insertion of $\Gamma \sim 1/N$ makes it of the same order as A. In the figure we show only three permutations B1, B2 and B3 out of the twelve possible. These are in fact the only diagrams we need to compute since one can show that each diagram has $\Bbb Z_2 \times \Bbb Z_2$ symmetry under exchange of momenta on each side of the exchange blob. Therefore the value of each diagram exclusively depends on which of the external momenta are above and which are below the blob corresponding to $\Gamma$ in the figure. In (A.11) we show an example of the B1 integrand.

\subsec{Double-Soft Limit}
Computing the Feynman diagrams of the previous section is a formidable task. Nevertheless based on the general argument of the previous section we know that the answer must take an extremely simple form \prop. In the rest of the section we will show that it is indeed the case and that the results are consistent with $c_i^{qf}=0$.

There is a particular kinematical regime in which the diagrams can be computed analytically. We will refer to it as the double soft limit \PRAKASH\ and it is defined as
\eqn\doublesoft{
\la O_{qf}(q_1=p)O_{qf}(q_2=0)O_{qf}(q_3=-p)O_{qf}(q_4=0)\ra ,
}
where, as explained above, the convention is to take the argument of the insertions to be the third component of the momentum, with the other ones being zero.\foot{The double soft configuration is sensitive on the specific way of taking the limit. Our prescription is to take the limit under the integrand. At the level of \ffinfs\ it corresponds to $\lim_{\varepsilon \to 0} F(p+\varepsilon, - \varepsilon,-p,0) $.} In the next section we will compute the integrals numerically for general collinear choices of $q_i$.  

The diagrams shown in \ABfig\ can be separated into two groups. For diagrams A, the result is different whether the soft momenta are adjacent, $A_{adj}$, or not, $A_{non-adj}$. For diagrams B the result depends on whether the two soft momenta are adjacent and on the same side of the exchange blob $B_{adj}$ or whether they are on opposite sides of the exchange blob $B_{non-adj}$.

Computation of the integrals involved in the diagrams is straightforward but the intermediate steps are cumbersome so we simply quote the final answers for each non-equivalent diagram. The dependence on $p$ is fixed by dimensional analysis to be $ \langle O_{qf}^4 \rangle_{double~soft} \sim 1/|p|$ so the exercise amounts to taking care of each pre-factor and checking if their sum saturates \prop~without a need for AdS$_4$ contact terms.

The diagrams where the soft momenta are not adjacent are the simplest ones to compute and their results are given by 
\eqn\Anadj{\eqalign{
A_{non-adj} &=- {1\over N} {\pi^5 \lambda(\pi \lambda + \sin \pi \lambda )\over 128 |p| \sin^2{\pi \lambda \over 2}} ,\cr
B_{non-adj} &= - {1\over N}{\pi^5 \lambda( 4 \tan {\pi \lambda \over 2}-\pi \lambda - \sin{\pi \lambda}) \over 256|p| \sin^2 {\pi \lambda \over 2}} ,
}}
which agrees with the results of \PRAKASH. The analytic calculation of adjacent diagrams is more involved and the answer is 
\eqn\Aadj{\eqalign{
A_{adj} &= - {1\over N} {\pi^5 \lambda \over 256|p|\sin^2 {\pi \lambda \over 2}}\left(\pi(1-2 \lambda)\lambda +\left(1-2\lambda+2 \lambda^2 ( \psi({1+\lambda \over 2}) - \psi({\lambda \over 2}))\right)\sin \pi \lambda \right),\cr
B_{adj} &=- {1 \over N} {\pi^5 \lambda^2 \over 128 |p| \sin^2 {\pi \lambda \over 2}}\left(\pi \lambda +\left(1+ \lambda \psi({\lambda \over 2}) - \lambda \psi({1+\lambda \over 2})\right)\sin \pi \lambda \right),
}}
where $\psi(x)=\Gamma'(x)/\Gamma(x)$ denotes the digamma function. Adding the diagrams and taking into account multiplicities of permutations, the result is\foot{The final result of \PRAKASH\ is different from ours.}
\eqn\doublesoft{\eqalign{
\la O_{qf}(p)O_{qf}(0)O_{qf}(-p)O_{qf}(0)\ra_{conn} &=2 A_{non-adj} + 4 A_{adj} + 8 B_{non-adj}+ 4 B_{adj} \cr
&= { \pi \lambda \over 2 N \sin \pi \lambda } \left(- {\pi^4 \over 2|p|}\right) \cr
&= {1 \over \tilde{N}} ~F_{ff}(p,0,-p,0),
}}
which is consistent with \prop. This is enough to claim that $c_1^{qf}+c_2^{qf}+c_3^{qf}=0$. To show that the answer is actually \identityQF\ with $c_i^{qf}=0$ we will study numerically arbitrary kinematics in the next section.

\subsec{General Collinear Kinematics}

\ifig\CheckForward{Connected scalar four-point function normalized as $-{|p| N\over \pi^4}\la \prod_i O_{qf}(q_i) \ra_{conn}$ (vertical axis) as a function of the 't Hooft coupling (horizontal axis). Blue dots: Numerical integration of Feynman diagrams A$+$B. Line: Expectation from \prop.} {\epsfxsize 2.9in\epsfbox{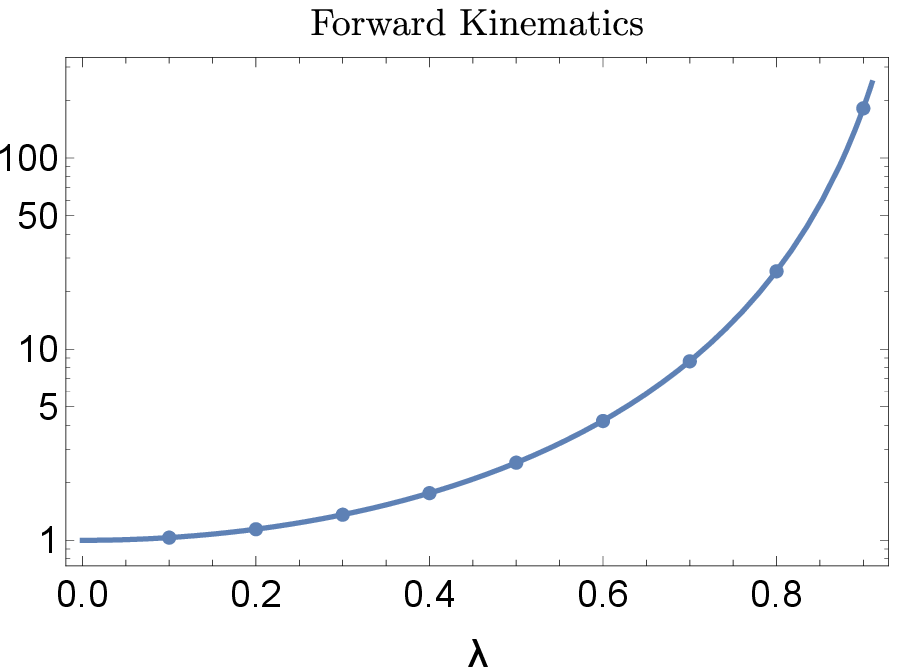}}

The double soft limit is very special since $q_1 \sim- q_3$ and $q_2 \sim q_4 \sim 0$. In this section we would like to test our proposal for arbitrary kinematics lacking any particular symmetry between external insertions, but still within the collinear configuration. Analytic computations look very hard in this case therefore we compute the correlator numerically.

An interesting configuration which is in some sense the opposite of the double soft limit (in which two momenta are hard and two are soft) is the forward limit in which all momenta are equal. We will numerically compute  
\eqn\forward{
\la O_{qf}(q_1=p)O_{qf}(q_2=-p)O_{qf}(q_3=p)O_{qf}(q_4=-p)\ra_{conn} ,
}
which we denote as forward limit in analogy to scattering. We show the result of this computation in \CheckForward\ where we compare the prediction with \prop, corresponding to the solid line.\foot{Again our prescription for taking the forward limit is to set the momenta to forward kinematics under the integrand.} We get a perfect agreement with our expectation.

\ifig\CheckGeneral{Connected scalar four-point function normalized as $-{N\over \pi^4}\la \prod_i O_{qf}(q_i) \ra_{conn}$ (vertical axis) as a function of the 't Hooft coupling (horizontal axis). Blue dots: Numerical integration of Feynman diagrams A$+$B. Line: Expectation from \prop.} {\epsfxsize 4in\epsfbox{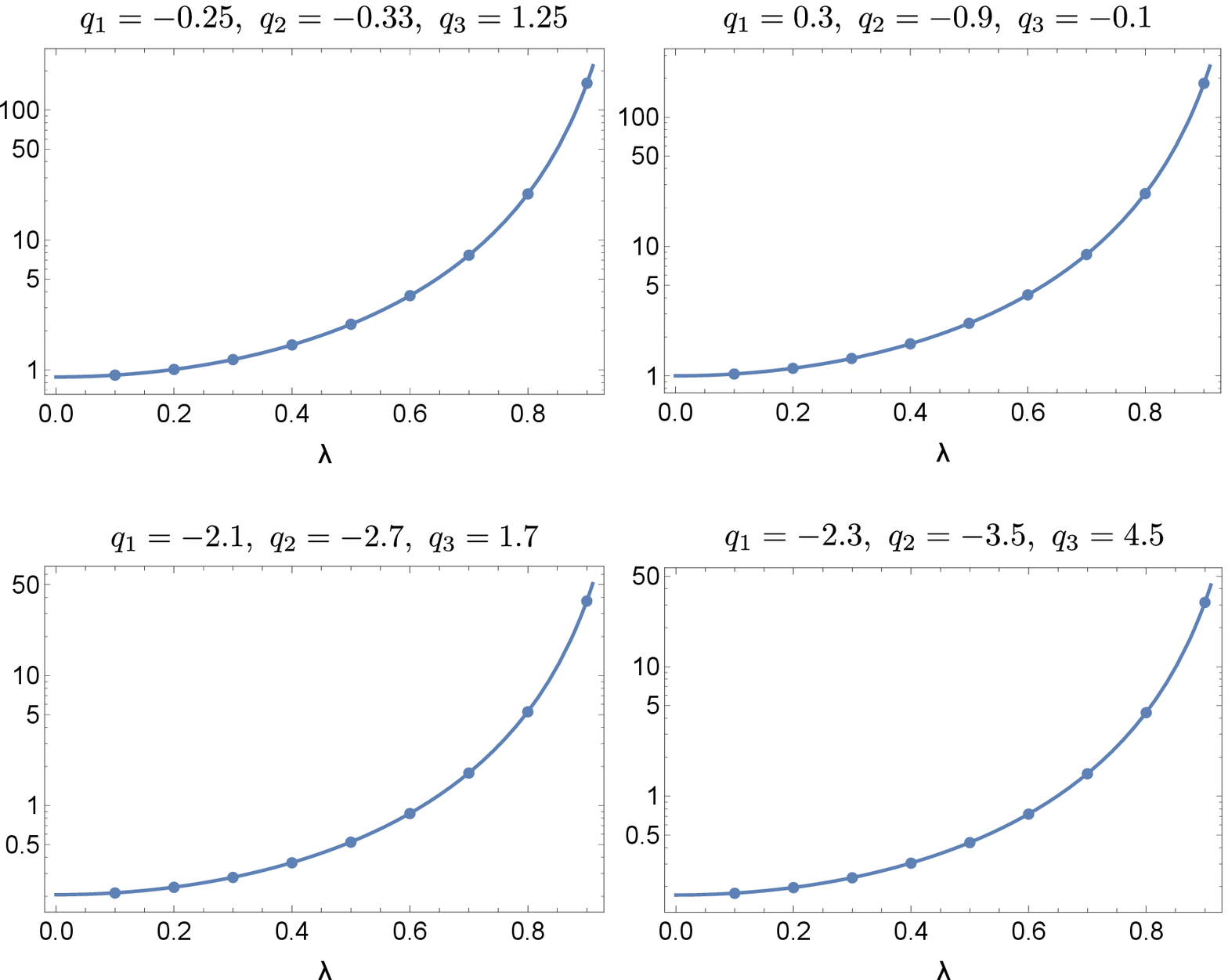}}

 To complete our check we evaluate the correlator in the most general collinear kinematics. The results are presented in \CheckGeneral\ and again we get the perfect agreement with the expected results. When comparing our prediction with the result of numerical integration we find on average agreement within $10^{-5}$\%. The estimated error of the numerical integration is $10^{-4}$\%. Needless to say, the simplicity of the final result is highly non-obvious at the intermediate steps which are pretty much intractable even in the collinear kinematics. 
 
\subsec{Quasi-Boson Theory}

Given the answer in the quasi-fermion theory we can compute the result in the quasi-boson theory using the Legendre transform \GurAriIS. Doing the Legendre transform is particularly simple in momentum space \PRAKASH, however, to connect with our prediction we will do the relevant computation in coordinate space. The relevant three-point function of scalar operators in the quasi-fermion theory is zero at separated points but it admits a conformal invariant contact term  \GurAriIS 
\eqn\contactterm{
\la O_{qf}(x_1) O_{qf}(x_2) O_{qf}(x_3) \ra = c \  \delta^{(3)}(x_{12}) \delta^{(3)}(x_{13}) .
}

\ifig\exchange{The vertices are given by \contactterm. The propagators are $\la \sigma(x) \sigma(0) \ra = {c_\sigma \over x^2}$. We integrate over the position of the vertices.} {\epsfxsize 1.8in\epsfbox{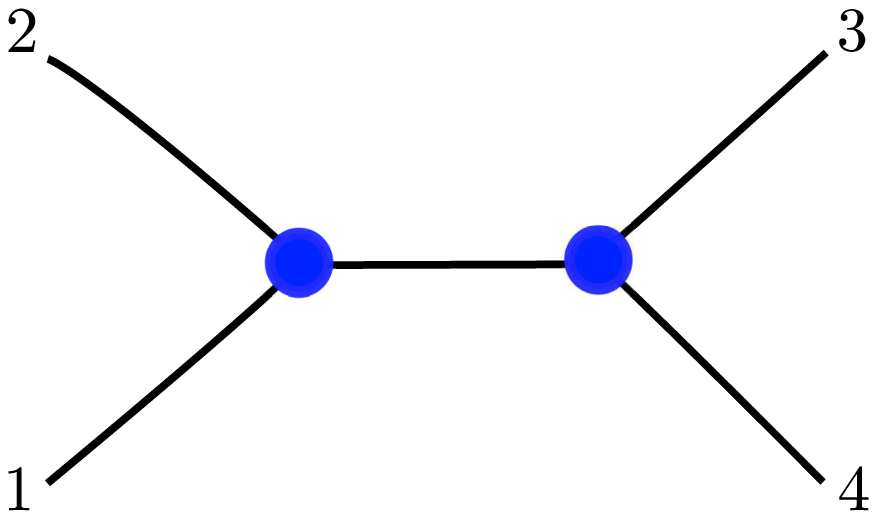}}

This contact term contributes to the Legendre transform through three exchange diagrams, see \exchange, which are related by permutations. The diagram takes the following form
\eqn\diagramcontr{\eqalign{
&c^2 c_\sigma^5\int d^3 y d^3 y' {1 \over (x_1 - y)^2 (x_2 - y)^2} {1 \over (y-y')^2} {1 \over (x_3 - y')^2 (x_4 - y')^2} \cr
&= {c^2 c_\sigma^5 \pi^3 \over |x_{34}| } \int d^3 y  {1 \over (x_{1} - y)^2 (x_2 - y)^2 |x_3 - y'| |x_4 - y'|} = c^2 c_\sigma^5 \pi^{7/2}  {\bar D_{1,1,{1 \over 2}, {1 \over 2}}(u,v) \over x_{13}^2 x_{24}^2} ,
} }
where we used standard identities \refs{\DolanUW,\DolanUT,\DolanIY} and the fact that $\la \sigma(x) \sigma(0) \ra = {c_\sigma \over x^2}$. Next we can use the results of \GurAriIS
\eqn\resultsnorm{\eqalign{
c &=- \tilde N {1 + \tilde \lambda_{qb}^2 \over \tilde \lambda_{qb}^3} \ , \cr
c_\sigma &=  {8 \over \pi^2} {\tilde \lambda_{qb}^2 \over 1 + \tilde \lambda_{qb}^2}  {1 \over  \tilde N }\ .
}}
To match with \quasiboson\ we rescale $\sigma \to {1 \over \sqrt{c_\sigma}} \sigma$, the result being $ c^2 c_\sigma^3 \pi^{7/2}   = {8 \over \pi^{5/2}} {1 \over  1 + \tilde \lambda_{qb}^2}  {1 \over  \tilde N }$ which coincides with the expected result \quasiboson\ modulo the absence of $-\tilde \lambda_{qb}^2$ in the numerator.

The reason for that is that the Legendre transform of the free fermion correlator has to be combined with ${1 \over \tilde N}  {8 \over \pi^{5/2}} \left( \bar D_{1,1,{1 \over 2}, {1 \over 2}}(u,v) + \bar D_{1,1,{1 \over 2}, {1 \over 2}}(v,u) + {1 \over u} \bar D_{1,1,{1 \over 2}, {1 \over 2}}({1 \over u},{v \over u}) \right)$ to produce the free boson correlator. It turns ${1 \over  1 + \tilde \lambda_{qb}^2}$ into $\left({1 \over  1 + \tilde \lambda_{qb}^2} - 1 \right) = - { \tilde \lambda_{qb}^2 \over  1 + \tilde \lambda_{qb}^2} $. It is a much simpler exercise to show it in momentum space which was done in \PRAKASH. 

The conclusion of this discussion is that the answer in the quasi-boson theory is given by \quasiboson\ with $c_{i}^{qb} = 0$.

\subsec{Critical Theories}

It is interesting to use the results that we obtained to compute the four-point functions in the critical $O(N)$ and Gross-Neveu models. The former corresponds to $\tilde \lambda_{qf} \to \infty$ and the latter to $\tilde \lambda_{qf} \to \infty$. 

For the critical $O(N)$ model our prediction is that the answer should be given by the free fermion answer. Indeed, this agrees with the results of \LeonhardtDU, where the answer was obtained based on crossing symmetry and direct evaluation of Feynman diagrams.  Similarly, in \PRAKASH\ the correlator was computed in the collinear kinematics $q^{\pm} = 0$ and again the answer was found to be the free fermion one. Our derivation confirms both of these results.

For the critical Gross-Neveu model we get
\eqn\predictionGN{
f_{GN}(u,v) =  f_{fb}(u,v) - {8 \over \pi^{5/2}}\left( \bar D_{1,1,{1 \over 2}, {1 \over 2}}(u,v) + \bar D_{1,1,{1 \over 2}, {1 \over 2}}(v,u) + {1 \over u} \bar D_{1,1,{1 \over 2}, {1 \over 2}}({1 \over u},{v \over u}) \right) ,
}
where recall that we normalized our operators such that the disconnected piece comes with the coefficient $1$. To our knowledge this prediction is new.

The connected four-point function in the planar limit contains both single and double trace operators. From \predictionGN\ one can compute for example the $1/N$ anomalous dimensions of double trace operators made up of $O_{qb}$, which matches the results in \GiombiRHM. On the other hand, the anomalous dimensions of double trace operators for the critical $O(N)$ model vanish to leading order in ${1 \over N}$ in three dimensions. 

\newsec{Conclusions}

The purpose of this paper was to compute four-point functions in large $N$ CFTs with slightly broken higher spin symmetry. Based on general arguments we found a three-parameter family of correlators \identityQF\ and \quasiboson . Our basic observation was that it is very easy to construct an ansatz for the four-point function which has the correct double discontinuity in every channel. The three free parameters are similar to subtraction terms in the scattering amplitudes dispersion relations. We then fixed the free parameters using the Schwinger-Dyson equations in the large $N$ Chern-Simons matter theories. The final result for the quasi-fermion correlator is \resultfourQF\ and \resultfourQB\ for the quasi-boson. These are the main results of the paper.

The characteristic feature of our result is that the answer is analytic in spin up to $J=0$. It would be interesting to understand if this feature persists for a generic spinning correlator. This sounds plausible because different four-point functions are related by higher spin symmetry (see appendix B). Moreover, since AdS contact diagrams survive in the flat space or bulk point limit \GaryAE,\MaldacenaIUA\ it would suggest existence of some type of sub-AdS locality in the higher spin gauge theories. Since bulk equations of motion respect higher spin symmetry this would seem to violate  the Coleman-Mandula theorem \ColemanAD. Therefore, we expect that maximal analyticity in spin is the property of every correlator in CFTs with weakly broken higher spin symmetry. If this is the case then no new input is needed to fix the four-point functions apart from already known three-point functions of single-trace operators. 

There are many interesting directions in which our work should be extended. It would be very interesting to compute four-point functions with spinning operators.\foot{For example theories of the type considered here were found to saturate the bounds coming from ANEC \refs{\HofmanAR\ChowdhuryVEL\CordovaZEJ-\MeltzerRTF}. This implies that certain four-point functions involving $j_2$ should indeed be almost free \refs{\ZhiboedovOPA-\MeltzerRTF}. It would be interesting to understand the implications of the energy correlators triviality (and their higher spin analogs) for correlation functions in this theory.} This would require a generalization of some of the CFT technology that we used for spinning correlators, see \refs{
\KarateevJGD,\KravchukDZD}. Since understanding of weakly broken higher spin symmetry is lacking we might hope that knowing generic four-point functions will give us some hints on how to think about these symmetries. Ideally, one would like to write some slightly broken higher spin invariants as was done in the unbroken case in \DidenkoTV.

Another direction is to generalize our arguments to higher-point functions. This would require a better understanding of the structure of the higher-point CFT correlators and their OPE structure. One may also hope to push further the diagrammatic computations in this direction. 

We have not explored the possibility of bootstrapping the correlators in momentum space using the higher spin Ward identities. The structure of the equations (see appendix B) suggests that it would be a natural place to study the correlators. One comment is that the momentum space computations are sensitive to contact terms which were not systematically understood in the large $N$ Chern-Simons matter theories to our knowledge.

All in all higher spin gauge theories duals seem to be a great playground for very general CFT ideas. Moreover, as we tried to demonstrate the correlators in these theories are much simpler than it seems at the intermediate steps. Therefore it should be possible to completely solve these theories in the planar limit. We do not know how to re-derive our results using the bulk theory. The simplicity of the final results suggests that there might be a simple argument for that. 

\newsec{Acknowledgments}

We are grateful to S. Giombi, J. Penedones, E. Skvortsov, A. Strominger, M. Taronna, R. Yacoby, X. Yin for useful discussions. The work of A.Z. is supported by NSF grant 1205550. We thank Simon Caron-Huot for pointing out that the structure $c_3$ was missing in the original version of the draft.

\appendix{A}{Schwinger-Dyson Approach: Details}
In this appendix we will give enough details about the computation done in section SEC to reproduce our results. The calculation is done in light cone gauge so we begin by recalling the definition of light cone coordinates $x^\pm = {x^1 \pm i x^2 \over \sqrt{2}}$, and with metric $ds^2=dx^+dx^- + (dx^3)^2$. Indices are lowered/raised as $p^\pm = p_{\mp}$. Following usual convention we denote the size of the components of a vector along 1-2 as $p_s^2 = 2 p^+ p^- =p_1^2 + p_2^2$. We denote the gamma matrices $\gamma^\mu$ with $\mu=+,-,3$, given by $\gamma^\pm=(\sigma^1 \pm i \sigma^2)/\sqrt{2}$ and $\gamma^3=\sigma^3$, where $\sigma^i$ are the Pauli matrices. We use upper case index $A=+,-,3,4$ if the identity $\gamma^4={\rm 1}$ is included.

 We work in a gauge $A_-=0$, in which the gauge boson propagator
$$
\la A^a_\mu(-p) A_\nu^b(q) \ra = G_{\nu\mu}(p) \delta^{ab} (2\pi)^3 \delta^3(q-p), 
$$
where $a$, $b$ label color indices, is given non perturbatively as 
\eqn\AA{
G_{+3}(p) =-G_{3+}(p)= {4 \pi i \lambda \over N} {1 \over p^+},
} 
with other components vanishing. We denote it in Feynman diagrams by a wavy line. On the other hand, the exact fermion propagator 
$$
\la \psi_i(p) \bar{\psi}^j(-q) \ra = S(p) \delta^j_i (2 \pi)^3 \delta^3(p-q)
$$ 
was computed in \GiombiKC 
\eqn\snonper{
S(p) = {-i \gamma^\mu p_\mu + \lambda p_s {\bf 1}+ i \lambda^2 \gamma^+ p^- \over p^2} .
}
We denote the full propagator in Feynman diagrams as a black oriented line. Finally we define the form factor which allows us to go from $O_{qf}$ to fermions
$$
\la O_{qf}(-q) \psi_i (k) \bar{\psi}^j(-p) \ra= V(p,q) \delta_i^j (2\pi)^3 \delta^3(q+p-k)
$$
where $V(p,q)$ is a $2\times2$ matrix. The momenta of $O_{qf}$ is along direction $3$ with value $q$, while $p$ and $k$ are arbitrary vectors. This correlator was obtained in \GurAriIS~by using Schwinger-Dyson equation techniques. If we parametrize the components of the matrix as $V(p,q)=v_A (p,q) \gamma^A$. We denote the form factors in Feynman diagrams by ``$\otimes$" and we take the momenta $q$ to be always incoming. 

\ifig\V{Vertex corresponding to the form factor between $O_{qf}(-q)$ and two fermions $\psi(k)$ and $\bar{\psi}(-p)$. For simplicity we omit color indices in the diagrams.} {\epsfxsize1.3in\epsfbox{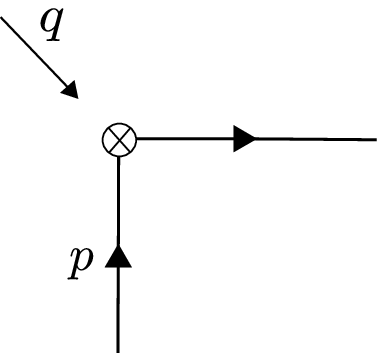}}

The solution found in \GurAriIS~is given by
\eqn\FFcomp{\eqalign{
v_I(p,q) &={\pi^{3/2} \lambda^{1/2} \over 2 N^{1/2} \tan^{1/2}{\pi \lambda \over 2}}   { 1 + e^{- 2 i \lambda \arctan{{2p_s \over q}}} \over 1 + e^{- \pi i \lambda {\rm sign}(q)} }  \cr
v_+(p,q)&={\pi^{3/2} \lambda^{1/2} \over 2 N^{1/2} \tan^{1/2}{\pi \lambda \over 2}}  {2\lambda p_+ \over p_s}{ 1-i\lambda {2p_s \over q}-(1+i\lambda {2p_s \over q}) e^{- 2 i \lambda \arctan{{2p_s \over q}}} \over \lambda {2p_s \over q} [1 + e^{- \pi i \lambda {\rm sign}(q)} ]},
}}
together with $v_-=v_3=0$. In \GurAriIS~a hard cut-off $\Lambda$ is used for integrals over $p_s$. In this section we present the formulas after taking $\Lambda \to \infty$ since the correlator we are interested in does not diverge. For some applications, for example computing the $O_{qf}$ two-point function the cut-off is necessary. In those cases $\Lambda$ can be restored by replacing 
$$
{\rm sgn}(q) \to {2 \over \pi } \arctan{2\Lambda \over q}. 
$$

Finally we need the $\Gamma$ vertex defined in \Gammadef. We use here the same labels of momenta as in \SDfig. It is defined by a Schwinger-Dyson equation \SDeq, whose solution was found in \Minwalla. We will use here the same parametrization as in \PRAKASH, namely
\eqn\GAB{
\Gamma_{AB}(k,q,r) = { 4\pi  i  \over N}    ~Tr(\gamma_A \gamma^P \gamma_B \gamma_Q) A_P^{~~Q}(k,q,r) ,
}
where we extracted the dependence on $N$ explicitly. This form has the advantage that only a few components of the matrix $A_P{}^Q(k,q,r)$ are non-vanishing. The expressions are still fairly lengthy so we extract a common prefactor and define $\tilde{A}$ as
\eqn\atilde{
A_P{}^Q(k,q,r) = { \lambda ~ \tilde{A}_P{}^Q \over 16 e^{- 2 i \lambda \arctan{{2r_s \over q}}} [1 + e^{- \pi i \lambda {\rm sgn}(q)} ](k^+-r^+)}.
}
Finally the values of $\tilde{A}$ entries that solve the Schwinger-Dyson equation in the case that $q$ points along $3$-direction are given by the following expressions. The only non-zero components are such that the first index is either identity $I$
\eqn\Amatrixidid{\eqalign{
\tilde{A}_I{}^I &=(e^{- \pi i \lambda {\rm sgn}(q)} -e^{- 2 i \lambda \arctan{{2k_s \over q}}})(1+e^{- 2 i \lambda \arctan{{2r_s \over q}}}) {2 k^+ \over q}\cr
&-(e^{- \pi i \lambda {\rm sgn}(q)} -e^{- 2 i \lambda \arctan{{2r_s \over q}}})(1+e^{- 2 i \lambda \arctan{{2k_s \over q}}}) {2 r^+ \over q},
}}
 \eqn\Amatrixidm{\eqalign{
\tilde{A}_I{}^- &=(e^{- 2 i \lambda \arctan{{2k_s \over q}}} -e^{- \pi i \lambda {\rm sgn}(q)})\Big((i {2\lambda r_s \over q}-1)+(1+i  {2\lambda r_s \over q})e^{- 2 i \lambda \arctan{{2r_s \over q}}}\Big){k^+ \over r^+}  \cr
&-\Big((1+i  {2\lambda r_s \over q})e^{- 2 i \lambda \arctan{{2r_s \over q}}} - (i {2 \lambda r_s \over q}-1)e^{- 2 i \lambda {\rm sgn}(q)} \Big)(1+e^{- 2 i \lambda \arctan{{2k_s \over q}}}), 
}}
or $+$ direction 
\eqn\Amatrixpid{\eqalign{
\tilde{A}_+{}^I  &=\Big(e^{- \pi i \lambda {\rm sign}(q)}-e^{- 2 i \lambda \arctan{{2r_s \over q}}} )((i {2\lambda k_s \over q}-1)+(1+i {2\lambda k_s \over q})e^{- 2 i \lambda \arctan{{2k_s \over q}}}\Big){r^+ \over k^+} \cr 
&+\Big((1+i {2 \lambda k_s \over q})e^{- 2 i \lambda \arctan{{2k_s \over q}}} - (i {2 \lambda k_s \over q}-1)e^{-\pi i \lambda {\rm sgn}(q)} \Big)\Big(1+e^{- 2 i \lambda \arctan{{2r_s \over q}}}\Big),
}}
\eqn\Amatrixpm{\eqalign{
\tilde{A}_+{}^-  &=\Big((1+i  {2 \lambda k_s \over q})e^{- 2 i \lambda \arctan{{2k_s \over q}}} - (i {2 \lambda k_s \over q}-1)e^{- \pi i \lambda {\rm sgn}(q)} \Big)\cr
&\times \Big((i {2 \lambda r_s \over q}-1)+(1+i  {2 \lambda r_s \over q})e^{- 2 i \lambda \arctan{{2r_s \over q}}}\Big){q \over 2 r^+}, \cr 
&+\Big((1+i  {2 \lambda r_s \over q})e^{- 2 i \lambda \arctan{{2r_s \over q}}} - (i {2 \lambda r_s \over q}-1)e^{- \pi i \lambda {\rm sgn}(q)} \Big)\cr
&\times   \Big((i {2 \lambda k_s \over q}-1)+(1+i {2 \lambda k_s \over q})e^{- 2 i \lambda \arctan{{2k_s \over q}}}\Big) {q \over 2 k^+}.
}}
If necessary we can restore the dependence on $\Lambda$ by using the same replacement as explained for the form factor. For computing scalar four-point function or for verifying the Schwinger-Dyson equation \SDeq~this is not necessary.

To conclude we will write explicitly some of the integrands shown in \ABfig\ to clarify notations. Out of the six diagrams A without the four-fermion vertex, the one shown in \ABfig\ is given by 
\eqn\IntegrandA{\eqalign{
A &= -N \int {d^3k\over (2 \pi)^3} {\rm Tr} [ S(k+q_1+q_2) V(k+q_1,q_2) S(k+q_1) V(k,q_1) S(k) \cr
&V(k-q_4,q_4)S(k-q_4)V(k+q_1+q_2,q_3)].
}}
Out of the twelve diagrams B with an insertion of the four-fermion vertex, the one labeled as B1 in \ABfig\ is given by
\eqn\IntegrandBone{\eqalign{
B1 &= -N^2 \int {d^3k\over (2 \pi)^3}{d^3r\over(2 \pi)^3} {\rm Tr} [ S(k+q_1+q_2) V(k+q_1,q_2) S(k+q_1) V(k,q_1) S(k) \cr
&\gamma^B S(r) V(r-q_4,q_4)S(r-q_4)V(r+q_1+q_2,q_3) S(r+q_1+q_2) \gamma^A \Gamma_{AB}(k,q,r)].
}}

\appendix{B}{Higher Spin Ward Identities}

In this appendix we sketch a strategy for finding four-point functions that involve operators with spin. We do not have any concrete results regarding this, but we hope this appendix could give some idea to the reader about the complications that one faces. Let us consider a simple example where one of the operators is the spin two current and let us focus on the quasi-fermion theory. In this case the relevant three-point couplings involve the three-point functions $\la j_2 O_{qf} j_{s} \ra$ of the scalar with two spinning operators. As described for example in \MaldacenaSF\ this involves two structures, the one of the free fermion and the parity-violating one which appears only in interacting theories. Due to the parity-violating structure we cannot simply write down an ansatz as before with the correct double discontinuity. However, we can try to analyze higher spin Ward identities. Indeed, knowing the result for the four-point function of scalar operators simplifies this task.

Based on the three-point functions (and our result for the four-point function of scalars) a natural ansatz for the four-point function is the following
\eqn\ansatzbh{
\la j_2 O_{qf} O_{qf} O_{qf} \ra = {1 \over \tilde N} {1 \over \sqrt{1 + \tilde \lambda_{qf}^2 }} \la j_2 O_{qf} O_{qf} O_{qf} \ra^{ff} +  {1 \over \tilde N} {\tilde \lambda_{qf} \over \sqrt{1 + \tilde \lambda_{qf}^2 }} \la j_2 O_{qf} O_{qf} O_{qf} \ra^{odd}  \ .
}
The free fermion piece has the correct three-point OPE coefficients with currents is crossing symmetric and is bounded in the Regge limit. It also correctly reproduces the double discontinuity that involves free fermion structure in $\la j_2 O_{qf} j_{s} \ra$. The odd piece should be responsible for the parity-violating part of the four-point function. We could also consider adding AdS contact interaction terms which would introduce non-analyticity in spin. Let us for simplicity assume that such terms are absent and instead try to write down the higher spin Ward identities.

Here we basically follow \MaldacenaSF. The transformation of the scalar under the simplest higher spin transformation takes the form\foot{Note that our normalization of the scalar operators is different from \MaldacenaSF. }
\eqn\transformscalar{
[Q, O_{qf}] = \pa_-^3 O_{qf} + {1 \over \sqrt{1 + \tilde \lambda_{qf}^2 }} \pa_- \left[ \pa_- j_{-3} - \pa_3 j_{--} \right] .
}

Using our results for the four-point functions of the scalars the higher spin Ward identity takes the schematic form
\eqn\resultWI{\eqalign{
&-{\tilde \lambda_{qf}^2 \over \sqrt{1 + \tilde \lambda_{qf}^2 }} \la j_2 O_{qf} O_{qf} O_{qf} \ra^{ff} + {\tilde \lambda_{qf} \over \sqrt{1 + \tilde \lambda_{qf}^2 }} \la j_2 O_{qf} O_{qf} O_{qf} \ra^{odd} + ... \cr
&=\alpha {\tilde \lambda_{qf} \over \sqrt{1 + \tilde \lambda_{qf}^2 }} \int \la O_{qf}  O_{qf}  \ra \left( {1 \over \sqrt{1 + \tilde \lambda_{qf}^2 }} \la j_2 O_{qf} O_{qf} O_{qf} \ra^{ff} + {\tilde \lambda_{qf} \over \sqrt{1 + \tilde \lambda_{qf}^2 }} \la j_2 O_{qf} O_{qf} O_{qf} \ra^{odd} + ...\right),
}}
where we used the fact that the free-fermion answer satisfies higher spin Ward identities; we suppressed all the detailed structure of the differential operators; the $...$ stands for terms where $j_2$ appears at points $2$, $3$ and $4$; $\alpha$ is some numerical constant. The structure of \resultWI\ looks consistent. Indeed, the dependence on $\lambda$ in the RHS matches the one in the LHS and it involves the change of parity (as it should). 

The equations \resultWI\ look quite complicated in coordinate space, they, however, drastically simplify in momentum space. Indeed, the equation \resultWI\ becomes algebraic. Considering these equations in momentum space have two possible difficulties. First, conformal invariance in momentum space is complicated \MaldacenaNZ, \BzowskiSZA. However, the explicit results for the free fermion four-point function that we encounter in this paper were not particularly hard, and it could be that it is everything that is needed. Second, we should worry about contact terms that were absent in the coordinate space analysis. It would be interesting to repeat the analysis of \MaldacenaSF\ in momentum space to clarify this point. In particular, higher spin symmetry should relate the standard physical contact terms that appear in the three-point function of stress tensors to other three-point couplings. However, it could happen that the structure of the higher spin Ward identities is easy in momentum space after all. We leave investigating these problems for the future.

\listrefs
\bye